\documentclass[aps,pre,color,psfig,epsf,notitlepage]{revtex4-1}
\usepackage{graphicx} 
\usepackage{color}
\usepackage{amsmath}
\usepackage{amsfonts}
\usepackage{amssymb}
\usepackage{enumitem}

\usepackage{babel}
\usepackage{txfonts}
 
\usepackage[position=top,caption=false]{subfig}

\usepackage{tikz}
\usepackage{tikz-3dplot}
\usepackage{circuitikz}
\usepackage{pgfplots}
\pgfplotsset{compat=newest}%
\usetikzlibrary{babel}

\usetikzlibrary{automata}
\usetikzlibrary{arrows}
\usetikzlibrary{positioning,calc}
\usetikzlibrary{graphs}
\usetikzlibrary{graphs.standard}
\usetikzlibrary{arrows,decorations.markings}
\usepackage{tkz-graph}
\usetikzlibrary{chains,fit,shapes}
\usetikzlibrary{calc}

\tikzset{every loop/.style={min distance=10mm,looseness=10}}
\tikzset{every state/.style={minimum size=2mm}}

\tikzset{snake it/.style={decorate, decoration=snake}}

\renewcommand{\vec}[1]{\boldsymbol{\mathrm{#1}}}%
\usepackage[T1]{fontenc}
\usepackage{lmodern}
\usepackage{textcomp}
\usepackage[babel=true]{microtype}
\usepackage{amsmath}
\usepackage{amsfonts}
\usepackage{amssymb}
\usepackage{mathrsfs}
\usepackage{mathtools}
\usepackage{bm}
\usepackage{esint}
\let\originalleft\left
\let\originalright\right
\renewcommand{\left}{\mathopen{}\mathclose\bgroup\originalleft}
\renewcommand{\right}{\aftergroup\egroup\originalright}

\def\bignicefrac#1#2{
    \left. {%
    \raise1.5ex\hbox{$\displaystyle#1$}%
    }%
    \kern-.5em%
    \middle/%
    \kern-.35em%
    {%
    \lower1.25ex\hbox{$\displaystyle#2$}%
    } \right.%
    }

\newcommand\boltz{{k_\text{B}}}

\makeatletter
\newcommand{\size}[1]{\bBigg@{#1}}
\makeatother

\usepackage{isomath}
\usepackage{graphicx}
\usepackage{tikz}
\usepackage{tikz-3dplot}
\usepackage{pgfplots}
\pgfplotsset{compat=newest}

\pgfplotsset{
    CI1/.style={
        legend image code/.code={%
            \node[anchor=center,fill=light-gray, opacity=0.25] at (0.3cm,0cm) {};
        }
    },
    CI2/.style={
        legend image code/.code={%
            \node[anchor=center,fill=yellow, opacity=0.5] at (0.3cm,0cm) {};
        }
    },
    CI3/.style={
        legend image code/.code={%
            \node[anchor=center,fill=blue, opacity=0.25] at (0.3cm,0cm) {};
        }
    },
    CI4/.style={
        legend image code/.code={%
            \node[anchor=center,fill=red, opacity=0.25] at (0.3cm,0cm) {};
        }
    },
}
\usepackage{readarray}
\usepackage{tkz-graph}
\GraphInit[vstyle = Simple]
\tikzset{
    VertexStyle/.style={shape=coordinate}
    }
\newcommand{\loopNarrowN}[1]{
    \Loop[
        dist = 2cm,
        dir = NO,
        style={
            out=67.5,
            in=112.5
            }
        ](#1)%
}
\newcommand{\loopNarrowS}[1]{
    \Loop[
        dist = 2cm,
        dir = SO,
        style={
            out=-67.5,
            in=-112.5
            }
        ](#1)%
}

\newcommand{\loopNarrowNW}[1]{
    \Loop[
        dist = 2cm,
        dir = SOWE,
        style={
            out=112.5,
            in=157.5
            }
        ](#1)%
}

\newcommand{\loopNarrowWSW}[1]{
    \Loop[
        dist = 2cm,
        dir = SOWE,
        style={
            out=225,
            in=180
            }
        ](#1)%
}
\newcommand{\loopNarrowENE}[1]{
    \Loop[
        dist = 2cm,
        dir = NOEA,
        style={
            out=0,
            in=45
            }
        ](#1)%
}

\usepackage{grffile}
\usepackage{siunitx}
\DeclareSIUnit \kT { \text { \ensuremath { \boltz T } } }
\usepackage{nicefrac}

\usepackage{titletoc}

\definecolor{light-gray}{gray}{0.75}
\definecolor{light-light-gray}{gray}{0.875}
\colorlet{dark-green}{green!70!black}
\usetikzlibrary{arrows.meta,bending,decorations.markings,intersections,automata,pgfplots.units,spy,circuits,circuits.ee.IEC,calc,math,patterns}
\usepgfplotslibrary{fillbetween,decorations.softclip,external}

\begin{document}

\title{Topological weight and structural diversity of polydisperse chromatin loop networks}

\author{Andrea Bonato$^1$, Enrico Carlon$^2$, Sergey Kitaev$^3$, Davide Marenduzzo$^4$, Enzo Orlandini$^5$}
\affiliation{$^1$ Department of Physics, University of Strathclyde, Glasgow G4 0NG, Scotland, United Kingdom \\
$^2$ Soft Matter and Biophysics, KU Leuven, Celestijnenlaan 200D, 3001 Leuven, Belgium\\
$^3$ Department of Mathematics and Statistics, University of Strathclyde, Glasgow, G1 1XH, United Kingdom \\
$^4$SUPA, School of Physics and Astronomy, The University of Edinburgh, Edinburgh, EH9 3FD, United Kingdom \\
$^5$Department of Physics and Astronomy, University of Padova and  INFN, Sezione Padova,
Via Marzolo 8, I-35131 Padova, Italy}

\begin{abstract}
Current biophysical models for transcriptionally active chromatin view this as a polymer with sticky sites, mimicking transcription units such as promoters and enhancers which interact via the binding of multivalent complexes of chromatin-binding proteins. It has been demonstrated that this model spontaneously leads to microphase separation, resulting in the formation of a network of loops with transcription units serving as anchors. Here, we demonstrate how to compute the topological weights of loop networks with an arbitrary 1D pattern of transcription units along the fibre (or `polydisperse' loop networks), finding an analogy with networks of electric resistors in parallel or in series. We also show how the BEST (de Bruijn, van Aardenne-Ehrenfest, Smith and Tutte) theorem in combinatorics can be used to find the combinatorial multiplicity of any class of loop networks. Our results can be used to compute the structural diversity, or Shannon entropy, of loop networks: we show that this quantity depends on the 1D patterning of transcription units along the chain, possibly providing a pathway to control transcriptional noise in eukaryotic genes. 
\end{abstract}

\maketitle

\section{Introduction}

Chromatin is a protein-DNA composite polymer that provides the building block of chromosomes, and it constitutes the form in which genomic information is stored in the nuclei of eukaryotic cells. Chromatin also provides the genomic substrate for fundamental intracellular processing of DNA, such as transcription and replication~\cite{Calladine1997,Alberts2014}. Long-standing observations suggest that the 3D structure of chromatin is functionally important for such processes: for instance, it is known that the 3D structure of a gene correlates with its transcriptional activity~\cite{Chiang2024}. 

Polymer models to determine chromatin structure in 3D are therefore important in this field, and several coarse-grained potentials have been developed to describe them (see, e.g.,~\cite{Rosa2008,Barbieri2012,Jost2014,DiPierro2016,Chiariello2016,Bianco2018}, and~\cite{Brackley2020,Chiang2021,Chiang2022} for a review of some of these). Typically, coarse-grained polymer models view chromatin as a copolymer, or heterogeneous polymer, where different beads may have different properties to reflect, among others, the local sequence and post-translational modifications in DNA-binding histone proteins, such as acetylation or methylation (see, e.g.,~\cite{Jost2014,Chiang2019,Chiang2024}).

\begin{figure}
\begin{center}
\includegraphics[scale=0.5]{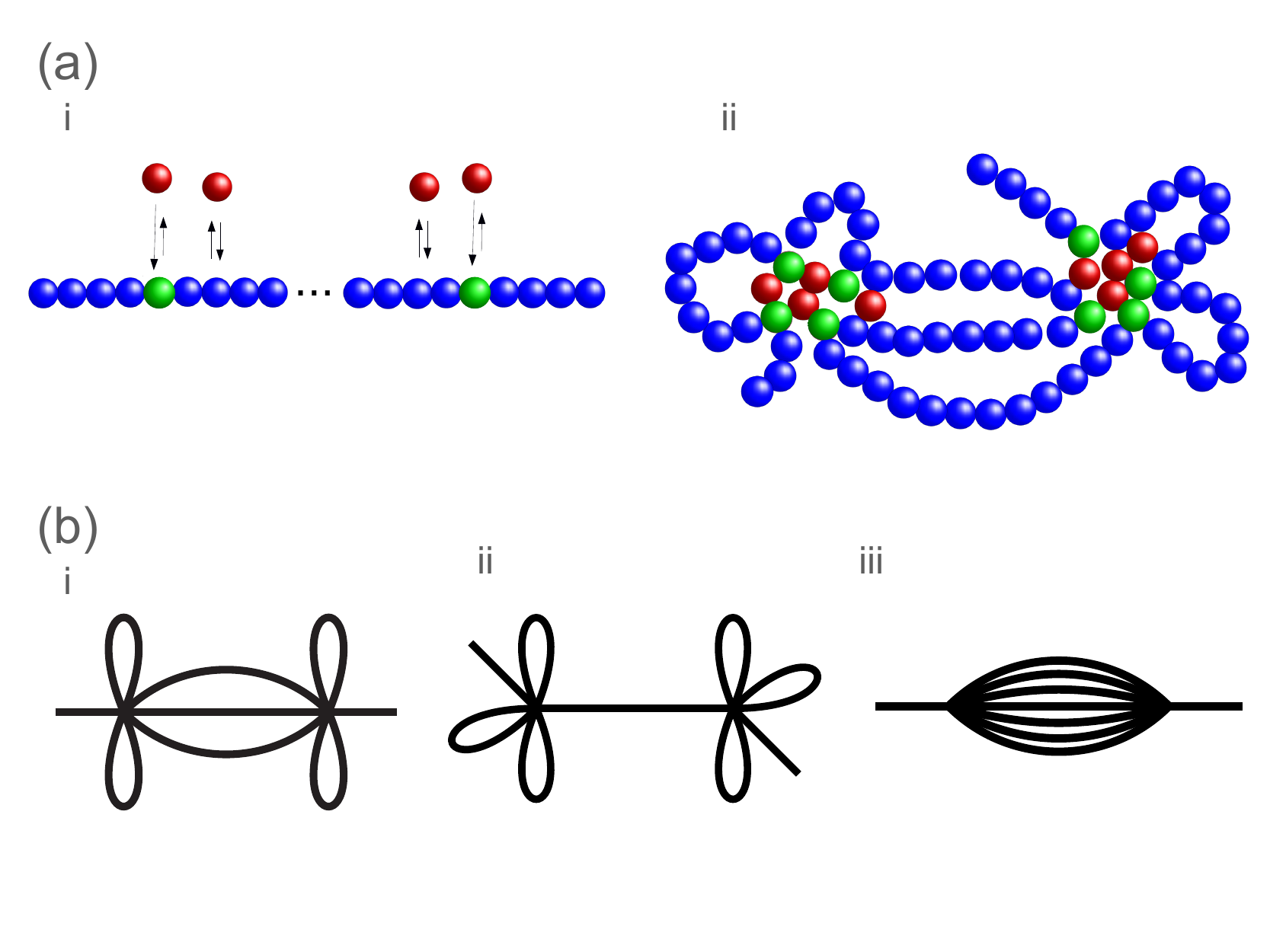}
\caption{(a) 
A chromatin fibre with $n$ TUs (green), interacting with chromatin-binding proteins
(red). The binding proteins have higher affinity for the TUs and lower for the 
rest of the fiber which induces a three-dimensional folding where TUs (effectively acting
as sticky sites) are the nodes of a complex polymer network. 
(b) A possible structure formed through bridging-induced phase separation~\cite{Brackley2016,short}. 
The structure is made of two clusters. 
(i) Loop network topology corresponding to the configuration in 
((a)ii). (ii) Rosette topology. (iii) Watermelon topology. }\label{fig1}
\end{center}
\end{figure}

A simple copolymer model for transcriptionally active chromatin~\cite{Marenduzzo2009,Brackley2021,short,Bonato2024}, which is relevant to our current work, views the fibre as a semiflexible polymer with interspersed ``transcription units'' (TUs, the green circles in Fig.~\ref{fig1}), representing open chromatin regions such as enhancers or promoters which have high 
affinity for multivalent chromatin-binding proteins associated with transcription -- such as RNA polymerases and 
transcription factors, or protein complexes including both of these~\cite{Brackley2016,Brackley2021}. 

The model in Fig.~\ref{fig1}A leads to the spontaneous formation of chromatin loop networks, through ``bridging-induced phase separation'' (or BIPS~\cite{Brackley2013,Brackley2016,Ryu2021}), a thermodynamic positive feedback loop between protein-chromatin binding and local clustering of binding sites, which may underlie the self-organisation of transcription factories in mammalian nuclei~\cite{Cook2018}. From the point of view of statistical and polymer physics, enumerating and finding the relative abundance of different types, or topologies, or loop networks is an interesting and non-trivial problem, and this is the topic of the present work.

Classical work on the statistical physics of polymer loop networks
considered the thermodynamic limit where the distance between sticky sites (or transcription units) was very large, in which case a perturbative renormalisation group calculation can provide the entropic exponents associated with each network~\cite{Duplantier1986,Duplantier1989}. In this limit, all networks with the same number of nodes and edges (or legs) emanating from each node have the same entropic exponent~\cite{Duplantier1989}. However, chromatin loop networks differ from these idealised ones, as the distance between sticky sites cannot be made arbitrarily large; rather, the thermodynamic limit of interest is the one in which the fibre becomes very long while the average density of transcription units along the fibre remains constant~\cite{short}. Recently, it was shown that the ``topological weight'' {(defined as the equilibrium partition function)}
of networks with the same entropic exponent, which determines their probability of occurrence, can be very different, and depends on the wiring, or topology, of the network~\cite{short,Bonato2024}: for instance, the rosette topology in Fig.~\ref{fig1}(b)ii is exceedingly more likely statistically than the watermelon topology in Fig.~\ref{fig1}(b)iii. 
In other words, the topological weight of the rosette topology is much larger. However, the calculations 
in~\cite{short,Bonato2024}
considered networks where the distance between TUs -- or sticky sites -- is fixed, and uniform: we call these networks ``monodisperse'', as all loops and junctions have the same length. In this work, we will instead consider the more general case where the distance between TUs is not fixed, but can take any value: we refer to the corresponding loop networks as ``polydisperse''.


As in~\cite{short,Bonato2024}, we consider two possible classes of chromatin loop networks. First, 'labelled' networks are those in which the TUs are numbered or labelled, so that they are all distinct from one another. This is often relevant in biological examples, where different TUs correspond to different regulatory elements, and it may be important in practice to distinguish networks with the same topology, but where different TUs participate in the clusters. Second, ``unlabelled'' networks are those where TUs are not numbered, or labelled, such that different configurations are topologically non-equivalent configurations of our chromatin fibre, which can be distinguished, for instance, via their network (or multigraph) adjacency matrix. 
Unlabelled networks are relevant when considering the relative likeliness of generic types of topologies, for instance, when examining the average over patterns genome-wide, where it does not make sense to label individual TUs. 




In the present work, our main goal is to find the topological weights and combinatorial multiplicities of networks such 
as those in Fig.~\ref{fig1}(b), and discuss how this can be potentially relevant to the biophysics of transcription. First, in Section II we discuss the representations of labelled and unlabelled polymer loop networks, showing that these can be efficiently provided by words, matrices and graphs. Second, in Section III we develop a theoretical framework to compute the topological weight of a generic labelled polydisperse loop network, under the approximation that the polymer is Gaussian (i.e., without accounting for self or mutual avoidance of the different polymer segments in the network). This shows that rosette-like topologies generally have a larger topological weight with respect to watermelon-like topologies; we shall also see that these weights resemble the effective impedance values arising in Kirchhoff networks of resistors. Third, in Section IV we consider the case of unlabelled, topologically inequivalent networks and show how the BEST theorem of combinatorics can be adapted to count the number of labelled networks corresponding to a given unlabelled network. We will find explicit formulas for general chain-like configurations (in the text) and for general three-cluster configurations (in Appendix A). Then, in Section V, we use the theoretical framework developed in Sections III and IV to compute the topological weights of all chromatin loop networks arising from a given arrangement of transcription units along the fibre. We will also find the Shannon entropy of the ensemble of these networks, which we shall refer to as the structural diversity of the underlying chromatin fibre: this measure quantifies the heterogeneity of chromatin folding at a gene, which is thought to correlate with transcriptional noise, or heterogeneity in gene expression~\cite{Chiang2024,Chiang2025}. Finally, Section VI contains our concluding remarks. 


\section{Representations of chromatin loop networks}

\begin{figure}

\centering{

\begin{tikzpicture}[scale=1.00]
    \SetGraphUnit{3}
    \Vertex{A}
    \EA(A){B}
    \WE[unit=1](A){in}
    \EA[unit=1](B){out}
    \Edge(A)(B)
    \Edge(in)(A)
    \Edge(B)(out)
    \loopNarrowN{A}
    \loopNarrowS{A}
    \loopNarrowN{B}
    \loopNarrowS{B}
    \foreach \angle in {45}
        {
        \tikzset{EdgeStyle/.append style = {bend left = \angle}}
        \Edge(A)(B)
        \Edge(B)(A)
        };

    \EA[unit=4](B){in2}
    \EA[unit=1](in2){C}
    \EA(C){D}
    \EA[unit=1](D){out2}
    \Edge(C)(D)
    \Edge(in2)(C)
    \Edge(C)(out2)
    \foreach \angle in {15,30,45}
        {
        \tikzset{EdgeStyle/.append style = {bend left = \angle}}
        \Edge(C)(D)
        \Edge(D)(C)
        };

    \node (note) at (-1.25,2) {{\fontsize{14}{0pt}\usefont{T1}{phv}{m}{n} (a)}};
    \node (note) at (6.75,2) {{\fontsize{14}{0pt}\usefont{T1}{phv}{m}{n} (c)}};

    \node (note) at (-0.5,0.25){{\textcolor{black}{a}}}; 
    \node (note) at (0.75,0.25) {{\textcolor{red}{1,2}}};
    \node (note) at (0.75,-0.25) {{\textcolor{red}{3,5}}};
    \node (note) at (2.25,0.25) {{\textcolor{red}{4,6}}};
    \node (note) at (2.25,-0.25) {{\textcolor{red}{7,8}}};
    \node (note) at (3.5,0.25){{\textcolor{black}{b}}}; 

    \node (note) at (8,0.75){{\textcolor{black}{a}}}; 
    \node (note) at (7.75,0.25) {{\textcolor{red}{1,3}}};
    \node (note) at (7.75,-0.25) {{\textcolor{red}{5,7}}};
    \node (note) at (11.25,0.25) {{\textcolor{red}{2,4}}};
    \node (note) at (11.25,-0.25) {{\textcolor{red}{6,8}}};
    \node (note) at (11,0.75){{\textcolor{black}{b}}}; 

    \node (note) at (-1.25,-2) {{\fontsize{14}{0pt}\usefont{T1}{phv}{m}{n} (b)}};
    \node (note) at (6.75,-2) {{\fontsize{14}{0pt}\usefont{T1}{phv}{m}{n} (d)}};

    \SO[unit=4](A){A3}
    \EA(A3){B3}
    \NOWE[unit=1](A3){in3}
    \SOEA[unit=1](B3){out3}
    \Edge(A3)(B3)
    \Edge(in3)(A3)
    \Edge(B3)(out3)
    \loopNarrowN{A3}
    \loopNarrowS{A3}
    \loopNarrowWSW{A3}
    \loopNarrowN{B3}
    \loopNarrowS{B3}
    \loopNarrowENE{B3}

    \EA[unit=4](B3){in4}
    \EA[unit=1](in4){A4}
    \NOEA[unit=1](A4){B4}
    \SOEA[unit=1](B4){C4}
    \EA[unit=1](C4){out4}
    \Edge(in4)(A4)
    \Edge(C4)(A4)
    \Edge(C4)(out4)
    \loopNarrowS{A4}
    \loopNarrowS{C4}
    \foreach \angle in {30}
        {
        \tikzset{EdgeStyle/.append style = {bend left = \angle}}
        \Edge(A4)(B4)
        \Edge(B4)(A4)
        };
    \foreach \angle in {-30}
        {
        \tikzset{EdgeStyle/.append style = {bend left = \angle}}
        \Edge(B4)(C4)
        \Edge(C4)(B4)
        };
    
    \node (note) at (-0.5,-3.75){{\textcolor{black}{a}}};
    \node (note) at (0.75,-3.75) {{\textcolor{red}{1,2}}};
    \node (note) at (0.75,-4.25) {{\textcolor{red}{3,4}}};
    \node (note) at (2.25,-3.75) {{\textcolor{red}{5,6}}};
    \node (note) at (2.25,-4.25) {{\textcolor{red}{7,8}}};
    \node (note) at (3.5,-4.25){{\textcolor{black}{b}}};

    \node (note) at (7.75,-3.25){{\textcolor{black}{a}}}; 
    \node (note) at (7.75,-3.75) {{\textcolor{red}{1,2}}};
    \node (note) at (7.65,-4.25) {{\textcolor{red}{5}}};    
    \node (note) at (9.0,-2.25){{\textcolor{black}{b}}}; 
    \node (note) at (9.0,-2.75) {{\textcolor{red}{3,6}}};
    
    \node (note) at (10.25,-3.75) {{\textcolor{red}{4,7}}};
    \node (note) at (10.35,-4.25) {{\textcolor{red}{8}}};
    \node (note) at (10.15,-3.25){{\textcolor{black}{c}}};

\end{tikzpicture}
}
\caption{(a-d) Examples of graph representations of chromatin loop networks. Labelled networks refer to these graphs with TU (nodes) labelled (here in red, in the order in which they are traversed). Unlabelled networks have no TU labelling but only cluster labelling (here in black).}\label{fig2}
\end{figure}
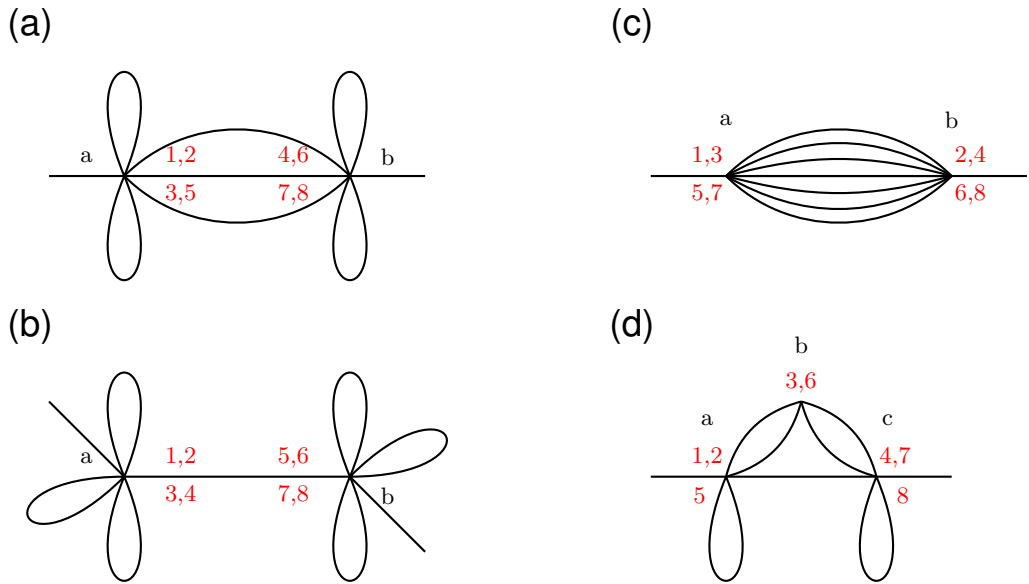

In this Section we discuss how to represent, or describe, labelled and unlabelled chromatin (or polymer) loop networks. One way that can always be used is via graphs, as shown in Fig.~\ref{fig2}(a-d). Labelled networks require numbering TUs (in red in Fig.~\ref{fig2}); removing the numbering gives a representation of the unlabelled version of the network.

Whilst graphs can be used to describe both labelled and unlabelled networks, it is also useful to discuss alternative representations: as we shall show later on, these alternative representations are helpful for combinatorial enumeration and for computing the topological weight of a loop network. 

Labelled networks -- or graphs with TU numbering, in red in Fig.~\ref{fig2}, in the order of traversal -- can be represented by strings, or words. In a string representation, letters refer to the cluster that a TU belongs to, while the letter position corresponds to the ordering of a TU along the chromatin fibre. For instance, the labelled loop networks in Fig.~\ref{fig2}(a-d) can respectively be represented by the following strings:
$$
{\rm (a):} \, aaababbb \qquad \qquad {\rm (b):} \, aaaabbbb \qquad \qquad {\rm (c):} \, abababab \qquad \qquad {\rm (d):} \, aabcabcc \, .
$$
The representation of suitable types of graphs (here labelled networks) with words is a more general topic in combinatorics: for more details, see~\cite{Kitaev2015}.

Unlabelled networks -- or graphs without TU numbering -- can be usefully represented by matrices that correspond to the adjacency matrix of the corresponding multigraph. These matrices are $n_c\times n_c$ matrices with $n_c$ the number of clusters. The diagonal entries are the number of loops at the different clusters, while the off-diagonal entries are the number of segments (or ties) joining the two corresponding clusters. Specifically, the matrix representations of the unlabelled version of the networks in Fig.~\ref{fig2} A-D are, respectively, given by:
\begin{equation}
	{\rm (a):} \,
	\begin{pmatrix}
		2 & 3 \\
		3 & 2 
	\end{pmatrix} \qquad \qquad 
    {\rm (b):} \,
    \begin{pmatrix}
        3 & 1 \\
        1 & 3 
    \end{pmatrix} \qquad \qquad
    {\rm (c):} \,
    \begin{pmatrix}
        0 & 7 \\
        7 & 0 
    \end{pmatrix} \qquad \qquad 
    {\rm (d):} \,
    \begin{pmatrix}
        1 & 2 & 1 \\
        2 & 0 & 2 \\
        1 & 2 & 1
    \end{pmatrix} \, .
\end{equation}
These matrix representations are useful for setting up the calculation of topological weights, of both monodisperse~\cite{Bonato2024} and polydisperse (see Section III) loop networks. {Obviously, for
the calculation of the topological weight of polydisperse networks one needs to provide, besides the topology, as given in terms of strings or matrices, the length of each segment as well.}

The networks in Fig.~\ref{fig2}(a-d) do not contain singletons (i.e., TUs which are not in a cluster, or joined to any other TU). However, involving singletons in the string or matrix representation is straightforward. For strings, they can be represented by a $0$ in the corresponding position: for instance, including a singleton TU in between TUs $4$ and $5$ in Fig.~\ref{fig2}(b) can be represented by the string $aaaa0bbbb$ 
Singletons do not affect the matrix representation, as they do not modify the inequivalent topology associated with them.

Finally, in what follows, it will sometimes be useful to refer to networks in which clusters are linearly arranged, one after the other, as chain-like configurations. Fig.~\ref{fig2}(a-c) provides simple examples of chain-like configurations: two-cluster networks are all chain-like. Instead, Fig.~\ref{fig2}(d) gives an example of a three-cluster configuration which is not chain-like; we will refer to types of networks like this as triangular configurations (these configurations are analysed in Appendix A).

\section{Topological weights of labelled polydisperse loop networks}

In this Section, we show how to compute the topological weight (or, equivalently, the partition function) of Gaussian polymer networks corresponding to any given chromatin loop network. The assumption of a Gaussian polymer means that our results hold for freely-jointed chains with a large number of monomers, but not for networks of self-avoiding and mutually-avoiding polymer segments~\cite{DeGennes1979}. Our calculation holds for generic polydisperse networks and hence generalises our previous theory in~\cite{Bonato2024}. We will use our generalised theory to show that rosette-like topologies with only or predominantly local loops are overwhelmingly more likely statistically than other types of topologies involving non-local loops (such as watermelon topologies). We will also use the results to compute the structural diversity (or Shannon entropy) of an ensemble of chromatin loop networks in Section V. Within this Section, we will also show that there is a nice and useful analogy between the calculation of the topological weight of a polymer loop network with the calculation of the
resulting resistance in a Kirchhoff resistor network.

In the following, we will base our theory on the case of monodisperse loops considered in~\cite{Bona2016}, showing how it can be modified to treat polydisperse loop networks. To start with, we define the topological weight of a given (labelled) graph ${\mathcal G}$ as its corresponding partition function, 
\begin{equation}\label{generaleqweight}
Z_{\mathcal G} = \int {d\mathbf x}_0 \ldots d{\mathbf x}_{n+1}\, \delta({\mathcal G})
\prod_{i=0}^{n} 
e^{-\frac{3\left({\mathbf x_{i+1}}-{\mathbf x_i}\right)^2}{2l_i\sigma}}.
\end{equation}
In Eq.~(\ref{generaleqweight}), $l_i$ denotes the distance between the $i$-th and the $(i+1)$-th TU, $\sigma$ is the bead size, and $\delta(\mathcal{G})$ is the product of Dirac delta functions that describes the topology of the network~\cite{Duplantier1986,Duplantier1989,Bonato2024}. In other words, $e^{-\frac{3\left({\mathbf x_{i+1}}-{\mathbf x_i}\right)^2} {2l_i\sigma}}$ can be thought of as the field-theoretical propagator from the $i$-th to the $(i+1)$-th TU. 

In the remainder of this Section, we will consider different types of graphs ${\mathcal G}$ and show how to compute Eq.~(\ref{generaleqweight}) in practice. 

\subsection{Topological weights of two-cluster configurations}

\begin{figure}

\centering{

\begin{tikzpicture}[scale=1.00]
    \SetGraphUnit{3}
    \Vertex{A}
    \EA(A){B}
    \NOWE[unit=1](A){in}
    \SOEA[unit=1](B){out}
    \Edge(A)(B)
    \Edge(in)(A)
    \Edge(B)(out)
    \loopNarrowN{A}
    \loopNarrowS{A}
    \loopNarrowWSW{A}
    \loopNarrowN{B}
    \loopNarrowS{B}
    \loopNarrowENE{B}

    \EA[unit=4](B){in2}
    \EA[unit=1](in2){C}
    \EA(C){D}
    \EA[unit=1](D){out2}
    \Edge(C)(D)
    \Edge(in2)(C)
    \Edge(C)(out2)
    \foreach \angle in {15,30,45}
        {
        \tikzset{EdgeStyle/.append style = {bend left = \angle}}
        \Edge(C)(D)
        \Edge(D)(C)
        };

    \node (note) at (-1.25,2) {{\fontsize{14}{0pt}\usefont{T1}{phv}{m}{n} (a)}};
    \node (note) at (6.75,2) {{\fontsize{14}{0pt}\usefont{T1}{phv}{m}{n} (b)}};

    \node (note) at (-1.25,1.25) {{\textcolor{red}{0}}};
    \node (note) at (0.75,0.25) {{\textcolor{red}{1,2}}};
    \node (note) at (0.75,-0.25) {{\textcolor{red}{3,4}}};
    \node (note) at (2.25,0.25) {{\textcolor{red}{5,6}}};
    \node (note) at (2.25,-0.25) {{\textcolor{red}{7,8}}};
    \node (note) at (4.25,-1.25) {{\textcolor{red}{9}}};
    \node (note) at (1.50,0.25) {{\textcolor{blue}{4}}};

    \node (note) at (6.75,0) {{\textcolor{red}{0}}};
    \node (note) at (7.75,0.25) {{\textcolor{red}{1,3}}};
    \node (note) at (7.75,-0.25) {{\textcolor{red}{5,7}}};
    \node (note) at (11.25,0.25) {{\textcolor{red}{2,4}}};
    \node (note) at (11.25,-0.25) {{\textcolor{red}{6,8}}};
    \node (note) at (12.25,0) {{\textcolor{red}{9}}};
    \node (note) at (9.5,0.85) {{\textcolor{blue}{$1,\cdots,7$}}};
    
\end{tikzpicture}
}
\caption{Loop network configurations, with TU and edge labelled (in red and blue respectively, in order of traversal). The labelling shown is used in the calculation of the topological weights of the (a) rosette and (b) watermelon topologies.}\label{topologicalweightRW}
\end{figure}
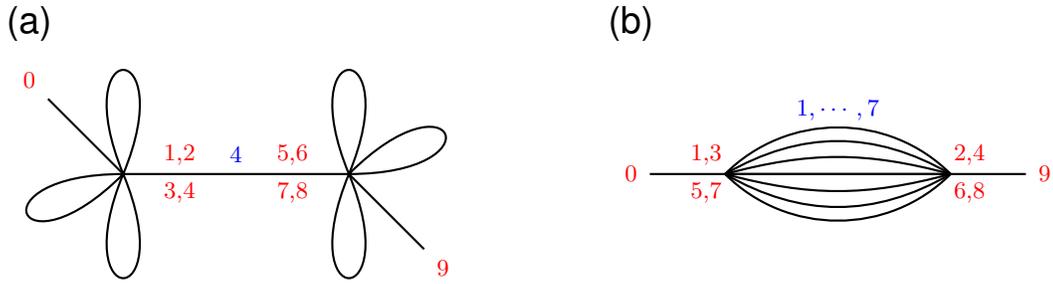

To begin with, we consider the case of two-cluster configurations, and for concreteness we specialise the calculation for the rosette and watermelon topologies considered in~\cite{Bonato2024}. The graph representations of these two topologies are shown in Fig.~\ref{fig2}A and B respectively: in these representations, nodes and edges are labelled (in red and blue respectively), in order of traversal, which is needed for our practical calculation. 

The term $\delta({\mathcal G})$ in Eq.~(\ref{generaleqweight}) is a product of Dirac $\delta$ functions that specify the topology of the network~\cite{Duplantier1989}: in the case of rosettes (${\mathcal G = \mathcal R}$, Fig.~\ref{topologicalweightRW}A) and watermelons (${\mathcal G = \mathcal W}$, Fig.~\ref{topologicalweightRW}B), 
\begin{eqnarray}
\delta({\mathcal R}) & = & \prod_{i=2,3,4}\delta({\mathbf x}_1-{\mathbf x}_i) 
\prod_{j=6,7,8} \delta({\mathbf x}_5-{\mathbf x}_j) \\ \nonumber
\delta({\mathcal W}) & = & \prod_{i=3,5,7}\delta({\mathbf x}_1-{\mathbf x}_i) \prod_{j=4,6,8}\delta({\mathbf x}_2-{\mathbf x}_j).
\end{eqnarray}

The topological weight of the rosette topology is then given by
\begin{equation}\label{polydisperserosetteweight}
Z_{\mathcal R} = \int d{\mathbf x}_0 \ldots d{\mathbf x}_9\,  
\left[\prod_{i=0}^{8} 
e^{-\frac{3\left({\mathbf x_{i+1}}-{\mathbf x_i}\right)^2}{2l_i\sigma}}\right]
\prod_{i=2,3,4}\delta({\mathbf x}_1-{\mathbf x}_i) 
\prod_{j=6,7,8} \delta({\mathbf x}_5-{\mathbf x}_j),
\end{equation}
where the TU labelling in the integral follows the one in Fig.~\ref{topologicalweightRW}A.

For ease of notation, we call
\begin{equation}
 \int d{\mathbf x}\, 
 e^{-\frac{3{\mathbf x}^2}{2l_i\sigma}} = \left(\frac{2\pi l_i\sigma}{3}\right)^{3/2}
 \equiv Z_i .
\label{def:Zi}
\end{equation}
By proceeding as in~\cite{Bonato2024}, we then obtain that the weight of a polydisperse rosette topology is 
\begin{eqnarray}\label{polydisperserosetteweight2}
Z_{\mathcal R} & = & Z_0 Z_8 \int d{\mathbf x}_1 d{\mathbf x}_5\,  
 e^{-\frac{3\left({\mathbf x_{1}}-{\mathbf x_5}\right)^2}{2l_4\sigma}} =  Z_0 Z_4 Z_9\,V,
\end{eqnarray}
where $V$ denotes the volume of the system.

Going through the same procedure for the watermelon topology (see Fig.~\ref{topologicalweightRW}B, and the associated choice of TU and edge labelling), we obtain that its topological weight is given by
\begin{eqnarray}\label{polydisperserosetteweight3}
Z_{\mathcal W} & = & Z_0 Z_8 \int d{\mathbf x}_1 d{\mathbf x}_2\,  
 \prod_{i=1}^{7}e^{-\frac{3\left({{\mathbf x}_2-{\mathbf x}_1}\right)^2}{2l_i\sigma}} 
 \\
 & = & Z_0 Z_8 \int d{\mathbf x}_1 d{\mathbf x}_2\,  
 e^{-\frac{3\left({{\mathbf x}_2-{\mathbf x}_1}\right)^2}{2L_{1,2,\ldots,7}\sigma}}
 = Z_0 Z_{1,2,\ldots,7}Z_8 V,
\end{eqnarray}
where we have defined:
\begin{eqnarray}\label{definitionL}
    \frac{1}{L_{1,2,\ldots,7}} & \equiv & \sum_{1=1}^{7} \frac{1}{l_i} = \frac{7}{H(l_1,l_2,\ldots, l_7)} \\ \nonumber
    Z_{1,2,\ldots,7} & \equiv & \left(\frac{2\pi L_{1,2,\ldots,7}\sigma}{3}\right)^{3/2},
\end{eqnarray}
where $H(l_1,l_2,\ldots,l_n)$ denotes the harmonic mean of $\{l_1,l_2,\ldots l_n\}$. It can be seen that the formula for $L_{1,2,\ldots,7}$ is the same as that for the resistance of a set of resistors in parallel with each other.

It is apparent that $Z_{\mathcal W}< Z_{\mathcal R}$. This follows from the inequality $\frac{1}{L_{1,2,\ldots,7}}=\sum_{i=1}^7 \frac{1}{l_i} > \frac{1}{l_4}$, which implies $L_{1,2,\ldots,7} < l_4$, and hence $Z_{1,2,\ldots,7}<Z_4$. Therefore, the topological weight of a polydisperse watermelon network is always smaller than that of the corresponding polydisperse rosette. 

The calculation shown above can be generalise in a straightforward way to compute the topological weight of a general two-cluster configuration with $n_l^{(1)}$ loops in the first cluster, $n_l^{(2)}$ loops in the second cluster, and $n_t$ ties (segments) connecting the two clusters. We will use this more general result when computing the Shannon entropy, or structural diversity, of a set of loop networks in Section V. 

\subsection{Topological weights of general polydisperse Gaussian loop networks}
\label{subsc:TWPGLN}

We now discuss how to calculate topological weights for a generic, not necessarily chain-like, chromatin loop network. 

\if{\begin{tikzpicture}[scale=1.00]
    \SetGraphUnit{3}
    \Vertex{A}
    \EA(A){B}
    \WE[unit=1](A){in}
    \EA[unit=1](B){out}
    \Edge(A)(B)
    \Edge(in)(A)
    \Edge(B)(out)
    \foreach \angle in {15,30,45}
        {
        \tikzset{EdgeStyle/.append style = {bend left = \angle}}
        \Edge(A)(B)
        \Edge(B)(A)
        };
\end{tikzpicture}}\fi

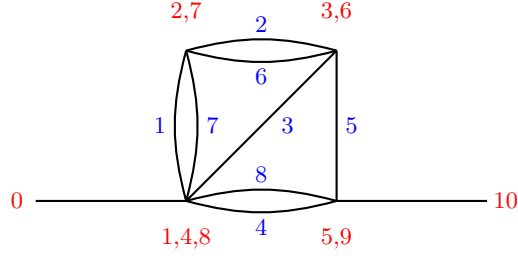
\begin{figure}
\begin{tikzpicture}[scale=1.00]
    \SetGraphUnit{3}
    \Vertex{A}
    \NO[unit=2](A){B}
    \WE[unit=2](A){in}
    \EA[unit=2](B){C}
    \SO[unit=2](C){D}
    \EA[unit=2](D){out}
    {
        \tikzset{EdgeStyle/.append style = {bend left = 15}}
        \Edge(A)(B)
        \Edge(B)(C)
        \Edge(A)(D) 
        \tikzset{EdgeStyle/.append style = {bend left = -15}}
        \Edge(A)(B)
        \Edge(B)(C)
        \Edge(A)(D)
    };
    \Edge(in)(A)
    \Edge(C)(A)
    \Edge(C)(D)
    \Edge(D)(out)
    \node (note) at (-2.25,0) {{\textcolor{red}{0}}};
    \node (note) at (0,-0.5) {{\textcolor{red}{1,4,8}}};
    \node (note) at (0,2.5) {{\textcolor{red}{2,7}}};
    \node (note) at (2,2.5) {{\textcolor{red}{3,6}}};
    \node (note) at (2,-0.5) {{\textcolor{red}{5,9}}};
    \node (note) at (4.25,0) {{\textcolor{red}{10}}};
    \node (note) at (-0.35,1.0) {{\textcolor{blue}{1}}};
    \node (note) at (0.35,1.0) {{\textcolor{blue}{7}}};
    \node (note) at (1,2.35) {{\textcolor{blue}{2}}};
    \node (note) at (1,1.65) {{\textcolor{blue}{6}}};
    \node (note) at (1,-0.35) {{\textcolor{blue}{4}}};
    \node (note) at (1,0.35) {{\textcolor{blue}{8}}}; 
    \node (note) at (1.35,1.0) {{\textcolor{blue}{3}}};
    \node (note) at (2.2,1.0) {{\textcolor{blue}{5}}};
    
\end{tikzpicture}
\caption{Loop network configuration with TU and edge labelling (in red and blue respectively) used for the calculation of the topological weights of a network with multiple clusters (here four).}\label{topologicalweightgenerictopology}
\end{figure}

For concreteness, we will consider the polydisperse generalisation of the same example network considered in~\cite{Bonato2024}, shown in Fig.~\ref{topologicalweightgenerictopology}: the labelling of TUs and edges, which we adopt in the calculation, is also shown. The associated topological weight of this labelled network is given by:
\begin{equation}\label{generictopologyweight}
Z_{\mathcal G} = \int d{\mathbf x}_0 \ldots d{\mathbf x}_{10}\,  
\left[\prod_{i=0}^{9} 
e^{-\frac{3\left({\mathbf x_{i+1}}-{\mathbf x_i}\right)^2}{2l_i\sigma}}\right]
\delta({\mathbf x}_1-{\mathbf x}_4)\,
\delta({\mathbf x}_1-{\mathbf x}_8)\,
\delta({\mathbf x}_2-{\mathbf x}_7)\,
\delta({\mathbf x}_3-{\mathbf x}_6)\,
\delta({\mathbf x}_5-{\mathbf x}_9).
\end{equation}
By applying the same methods used for the two-cluster configuration, this weight can also be written as
\begin{eqnarray}\label{generictopologyweight2}
Z_{\mathcal G} & = & Z_0 Z_9 \int d{\mathbf x}_1 d{\mathbf x}_{2} d{\mathbf x}_3 d{\mathbf x}_5\, e^{-\frac{3}{2\sigma} f({\mathbf x}_1, {\mathbf x}_2, {\mathbf x}_3, {\mathbf x}_5)}, 
\end{eqnarray}
where
\begin{eqnarray}
\label{equationf}
f({\mathbf x}_1, {\mathbf x}_2, {\mathbf x}_3, {\mathbf x}_5) & = &  
\left(\frac{1}{l_1}+\frac{1}{l_7}\right)\left({\mathbf x}_2-{\mathbf x}_1\right)^2+\left(\frac{1}{l_2}+\frac{1}{l_6}\right)\left({\mathbf x}_3-{\mathbf x}_2\right)^2+\frac{1}{l_3}\left({\mathbf x}_3-{\mathbf x}_1\right)^2 \\ \nonumber
& + & 
\frac{1}{l_5} {\left({\mathbf x}_5-{\mathbf x}_3\right)^2}+\left(\frac{1}{l_4}+\frac{1}{l_8}\right)\left({\mathbf x}_5-{\mathbf x}_1\right)^2 .
\end{eqnarray}

Similarly to the case of the monodisperse network~\cite{Bonato2024}, it is useful to introduce the following matrix, 
\begin{equation}
	A(\mathcal{G})=
	\begin{pmatrix}
		0 & \frac{1}{l_1}+\frac{1}{l_7} & \frac{1}{l_3} & \frac{1}{l_4}+\frac{1}{l_8} \\
		\frac{1}{l_1}+\frac{1}{l_7} & 0 & \frac{1}{l_2}+\frac{1}{l_6} & 0 \\
		\frac{1}{l_3} & \frac{1}{l_2}+\frac{1}{l_6} & 0 & \frac{1}{l_5} \\
        \frac{1}{l_4}+\frac{1}{l_8} & 0 & \frac{1}{l_5} & 0
	\end{pmatrix},
\end{equation}
which, if we define, in analogy with Eq.~(\ref{definitionL}), $L^{-1}_{i_1, \cdots, i_k}\equiv \sum_{j=1}^{k} \frac{1}{l_{i_j}}=\frac{k}{H_{i_1,\cdots, i_k}}$, can be written in a more compact form as 
\begin{equation}
	A(\mathcal{G})=
	\begin{pmatrix}
		0 & L^{-1}_{1, 7} & l^{-1}_{3} & L^{-1}_{4,8} \\
		L^{-1}_{1,7} & 0 & L^{-1}_{2,6} & 0 \\
		l^{-1}_{3} & L^{-1}_{2,6} & 0 & l^{-1}_{5} \\
        L^{-1}_{4,8} & 0 & l^{-1}_{5} & 0
	\end{pmatrix}.
\end{equation}
For the case where all $l_i$'s are equal, considered in~\cite{Bonato2024}, this matrix equals the adjacency 
matrix of the multigraph corresponding to $\mathcal{G}$.
Starting from $A(\mathcal{G})$, we define the related matrix
\begin{equation}
	B(\mathcal{G})=
	\begin{pmatrix}
		\Delta_1 & -L^{-1}_{1, 7} & -l^{-1}_{3} & -L^{-1}_{4,8} \\
		-L^{-1}_{1,7} & \Delta_2 & -L^{-1}_{2,6} & 0 \\
		-l^{-1}_{3} & -L^{-1}_{2,6} & \Delta_3 & -l^{-1}_{5} \\
        -L^{-1}_{4,8} & 0 & -l^{-1}_{5} & \Delta_4
	\end{pmatrix},
\end{equation}
where $\Delta_i$ equals the sum of the off-diagonal elements of the $i$-th row of the original matrix $A(\mathcal{G})$, so in our case
\begin{eqnarray}
    \Delta_1 & = & L^{-1}_{1, 7} + l^{-1}_{3} + L^{-1}_{4,8} \\ \nonumber
    \Delta_2 & = & L^{-1}_{1,7} + L^{-1}_{2,6} \\ \nonumber
    \Delta_3 & = & l^{-1}_{3} + L^{-1}_{2,6} + l^{-1}_{5} \\ \nonumber
    \Delta_4 & = & L^{-1}_{4,8} + l^{-1}_{5}.
\end{eqnarray}
This matrix can be used to write Eq.~(\ref{equationf}) as  
\begin{equation}
f({\mathbf x})={\mathbf{x}}^T B({\mathcal G}) {\mathbf{x}},
\end{equation}
where $\mathbf{x}^T=({\mathbf x}_1, {\mathbf x}_2,{\mathbf x}_3,{\mathbf x}_5)$. 

The topological weight of $\mathcal{G}$ can then be found by fixing  the position of the centre of mass of one of the clusters, and integrating over it, which gives
\begin{eqnarray}
Z_{\mathcal G} & = & Z_0 Z_9 {\bar{Z}}\,V\det(B'(\mathcal{G}))^{-3/2} 
\label{eq:ZG-det} \\ \nonumber
\bar{Z} & = & \left(\frac{2\pi \sigma}{3}\right)^{3/2 \, {\rm rk}(B'(\mathcal{G}))} 
\end{eqnarray}
where $B'(\mathcal{G})$ is any of the matrices obtained by removing the $i$-th row and column of $B(\mathcal{G})$ (this corresponds to integrating over the position of the centre of mass of the $i$-th cluster), and ${\rm rk}$ denotes the rank of a matrix (or, in this case equivalently, its dimension, which is $3$ in this example). 

We note that an alternative procedure to using the matrix $B'({\mathcal G})$ is to change variable to relative distances, by defining ${\mbox{\boldmath${\xi}$}}_i\equiv{\mathbf x}_{i+1}-{\mathbf x}_i$. In that case, instead of $B'({\mathcal G})$, the $3\times 3$ matrix of which we need to compute the determinant becomes
\begin{equation}
	\begin{pmatrix}
		L^{-1}_{1, 7}+l^{-1}_3+L^{-1}_{4,8} & l^{-1}_{3}+L^{-1}_{4,8} & L^{-1}_{4,8} \\
		l^{-1}_{3}+L^{-1}_{4,8} & L^{-1}_{2,6}+l^{-1}_3+L^{-1}_{4,8} & L^{-1}_{4,8} \\
		L^{-1}_{4,8} & L^{-1}_{4,8} & l^{-1}_5+L^{-1}_{4,8} 
	\end{pmatrix}.
\label{def_matrix_translation}
\end{equation}
It can be checked that the determinant of this matrix is the same as that of $B'({\mathcal G})$.

By applying the procedure outlined in this Section to a generic graph ${\mathcal G}$ with $n$ TUs and $k$ clusters, its topological weight can be computed, for example, by building the matrices $A({\mathcal G})$, $B({\mathcal G})$ and $B'({\mathcal G})$ as done in the example above. 

\subsection{Analogy with Kirchhoff resistor networks}
\label{sec:kirchhoff}


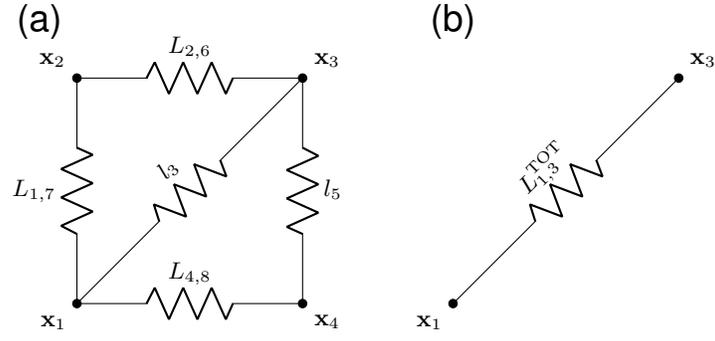
\begin{figure}
\begin{center}
\begin{circuitikz}[american voltages]
\draw

  (0,0) to [american resistor, l=$L_{1,7}$] (0,3)
  (0,0) to [american resistor, l=$L_{4,8}$] (3,0)
  (0,3) to [american resistor, l=$L_{2,6}$] (3,3)
  (3,3) to [american resistor, l=$l_5$] (3,0)
  (0,0) to [american resistor, l=$l_3$] (3,3)

  (5,0) to [american resistor, l=$L^{\rm TOT}_{1,3}$] (8,3);

  \node[circ](1) at (0,0)[label={[label distance=0.cm]245:${\mathbf x}_1$}]{};
  \node[circ](2) at (0,3)[label={[label distance=0.cm]135:${\mathbf x}_2$}]{};
  \node[circ](3) at (3,3)[label={[label distance=0.cm]45:${\mathbf x}_3$}]{};
  \node[circ](4) at (3,0)[label={[label distance=0.cm]315:${\mathbf x}_4$}]{};

  \node[circ](5) at (5,0)[label={[label distance=0.cm]245:${\mathbf x}_1$}]{};
  \node[circ](6) at (8,3)[label={[label distance=0.cm]45:${\mathbf x}_3$}]{};

   \node (note) at (-0.5,3.75) {{\fontsize{14}{0pt}\usefont{T1}{phv}{m}{n} (a)}};
   \node (note) at (5.0,3.75) {{\fontsize{14}{0pt}\usefont{T1}{phv}{m}{n} (b)}};
  
\end{circuitikz}
\end{center}
 \caption{(a) Resistor network representation of the polymer network of 
    Fig.~\ref{topologicalweightgenerictopology}. (b) Integrating out the node positions 
    $\vec{x}_2$ and $\vec{x}_5$ leaves a two nodes network with total resistance 
    $L^\text{TOT}_{1,3}$ given by Eq.~\eqref{eq:Ltot}.}
\label{fig-KH}
\end{figure}

While the calculation of the topological weight via the matrix determinant, as
done in the previous Section, is quite general, it is possible to perform the same calculation for simple topologies using an analogy with resistor networks. We illustrate this alternative 
method, which is rather simple, for the network in Fig.~\ref{topologicalweightgenerictopology}. 
The approach consists of integrating out the node positions one by one, and it is based on the following equality:
\begin{equation}
    \int d\vec{x} \,\, e^{-\frac{3}{2\sigma} \left[ \frac{(\vec{x}_i - \vec{x})^2}{l_i} + 
    \frac{(\vec{x} - \vec{x}_j)^2}{l_j}\right]}  =
    \left( \frac{2\pi \sigma}{3} \, \frac{l_i l_j}{l_i+l_j}\right)^{3/2} 
    e^{- \frac{3}{2\sigma} \frac{(\vec{x}_i - \vec{x}_j)^2}{l_i+l_j}} =
    \left( \frac{2\pi \sigma L_{i,j}}{3} \right)^{3/2} 
    e^{- \frac{3}{2\sigma} \frac{(\vec{x}_i - \vec{x}_j)^2}{l_i+l_j}},
\label{eq:series}
\end{equation}
where we used the same notation as in the previous Section.
The branches of the network can be seen as resistors with resistance $l_i$. 
Integrating over the position of the node $\vec{x}$ leads to a propagator with resistance given by the sum of resistances $l_i+l_j$ and weight that contains $L_{i,j} \equiv 
l_i l_j/(l_i+l_j)$, which is the formula for resistors in parallel.
A simple application of this rule to the network of Fig.~\ref{topologicalweightgenerictopology}, using the same notation as in the previous Section leads to
\begin{equation}
    Z_{\cal G} = V 
    Z_0 Z_9 
    \left( \frac{2\pi \sigma}{3} \frac{L_{1,7} L_{2,6}}{L_{1,7} + L_{2,6}} \right)^{3/2}
    \left( \frac{2\pi \sigma}{3} \frac{L_{4,8} l_5}{L_{4,8} + l_5} \right)^{3/2}
    \left( \frac{2\pi \sigma}{3} L^\text{TOT}_{1,3}  \right)^{3/2},
\label{eq:ZG-kirch}
\end{equation}
where the last three factors are obtained by integrating first the positions $\vec{x}_2$
and $\vec{x}_5$ and then, subsequently, $\vec{x}_1$ and $\vec{x}_3$, a procedure illustrated in Fig.~\ref{fig-KH}. The total resistance between nodes 1 and 3 is obtained by setting the resistors of the three branches in parallel
\begin{equation}
    \frac{1}{L^\text{TOT}_{1,3}} = \frac{1}{L_{1,7} + L_{2,6}} + \frac{1}{L_{4,8}+l_5} + \frac{1}{l_3}.
\label{eq:Ltot}
\end{equation}
Inserting \eqref{eq:Ltot} into \eqref{eq:ZG-kirch} we get
\begin{equation}
    Z_{\cal G} = V \left(\frac{2\pi \sigma}{3} \right)^{9/2} Z_0 Z_9 \,
    \left[ \frac{L_{1,7} L_{2,6} L_{4,8} l_3 l_5}
    {(L_{4,8} + l_5) l_3 + (L_{1,7} + L_{2,6})(L_{4,8} + l_5) + (L_{1,7} + L_{2,6}) l_3}
    \right]^{3/2}.
\label{ZG-via-kirchhoff}
\end{equation}
The calculation of the topological weight via the determinant of the
matrix $B'$ of Section~\ref{subsc:TWPGLN} or, equivalently, of the matrix \eqref{def_matrix_translation} 
gives the same result as Eq.~\eqref{ZG-via-kirchhoff}. The calculation via equivalent resistors network appears to be somewhat simpler. The determinant contains several positive and negative  terms, which partially cancel out and give a final, shorter expression of the weight. The resistor
network calculation yields directly a simpler and factorized structure originating from the
one-by-one integration of the nodes using \eqref{eq:series}.

We note a difference between the above approach and
the laws of electric circuits in which one determines the total
resistance for a voltage applied at two specific nodes of the resistor's network. In the
computation of the topological weight of a polymer network, one integrates over all nodes 
positions without any specific reference node. In the computation of the electric current through 
a resistors network, instead, it makes a difference whether the voltage is applied, say, to nodes 
$2$ and $5$, or to nodes $1$ and $5$. For instance, the rules of resistors in series/parallel are not applicable 
when the voltage difference is applied at nodes $2$ and $5$ of the network in Fig.~\ref{fig-KH}(a), 
but one can use these rules when the voltage difference is applied across nodes $1$ and $3$. In our 
calculation of $Z_{\cal G}$, we can choose which node position to integrate first, to make use of the rules of resistor networks in series/parallel given in Eq.~\eqref{eq:series}. 
Although in some complex networks we will not be able to reduce the network to a 
single resistor as Fig.~\ref{fig-KH}(b), one can apply the reduction formula 
\eqref{eq:series} to eliminate as many nodes as possible to finally remain with 
an ``irreducible'' network, to which the determinant formula \eqref{eq:ZG-det} can be 
always applied. According to this definition all networks in Fig.~\ref{fig2} are fully 
reducible to a single resistor.

\subsection{Expanded networks}
\label{sec:expanded}

\begin{figure}
    \begin{center}
\begin{tikzpicture}    \SetGraphUnit{3}
    \Vertex{A}
    \EA(A){B}
    \NOWE[unit=1](A){in}
    \SOWE[unit=1](A){out}
    \foreach \angle in {30}
        {
        \tikzset{EdgeStyle/.append style = {bend left = \angle}}
        \Edge(A)(B)
        \Edge(B)(A)
        };
    \Edge(in)(A)
    \Edge(A)(out)
    \loopNarrowN{A}
    \loopNarrowS{A}
    \loopNarrowN{B}
    \loopNarrowS{B}

    \draw [thick] (5.75,0.75) -- (7.0,0.25) arc (180:0:0.35 and 1.35) -- (7.7,0.25) arc(130:50:1.55cm) -- (9.7,0.25) arc (168:0:0.35 and 1.35); 
    \draw [thick] (5.75,-0.75) -- (7.0,-0.25) arc (-180:0:0.35 and 1.35) -- (7.7,-0.25) arc(-130:-50:1.55cm) -- (9.7,-0.25) arc (-168:0:0.35 and 1.35); 

    \draw[blue, thick, dotted] (7.0,0.25) -- (7.7,0.25); 
    \draw[blue, thick, dotted] (9.7,0.25) -- (10.4,0.0); 
    \draw[blue, thick, dotted] (10.4,0.0) -- (9.75,-0.25);
    \draw[blue, thick, dotted] (7.0,-0.25) -- (7.7,-0.25); 
    \draw[blue, thick, dotted] (7.7,0.25) -- (7.7,-0.25);
   
    \node (note) at (-1.0,0.25) {{\textcolor{red}{1,2}}};
    \node (note) at (-1.0,-0.25) {{\textcolor{red}{6,7}}};
    \node (note) at (3.5,0) {{\textcolor{red}{3,4,5}}};

    \node (note) at (6.85,0.6) {{\textcolor{red}{1}}};
    \node (note) at (7.85,0.6) {{\textcolor{red}{2}}};
    \node (note) at (9.5,0.6) {{\textcolor{red}{3}}};
    \node (note) at (10.6,0) {{\textcolor{red}{4}}};
    \node (note) at (9.5,-0.6) {{\textcolor{red}{5}}};
    \node (note) at (7.85,-0.6) {{\textcolor{red}{6}}}; 
    \node (note) at (6.85,-0.5) {{\textcolor{red}{7}}};

    \node (note) at (-1.25,2) {{\fontsize{14}{0pt}\usefont{T1}{phv}{m}{n} (a)}};
    \node (note) at (6.75,2) {{\fontsize{14}{0pt}\usefont{T1}{phv}{m}{n} (b)}};

\end{tikzpicture}

    \end{center}
    \caption{Examples of (a) tight and (b) expanded network. TUs are labelled 
    in red, whereas dotted blue lines indicate nodes bound by harmonic ``springs'' 
    which have an associated weight given in Eq.~\eqref{def:ps}, with an equivalent 
    ``resistance'' $K$, denoting the inverse spring stiffness.
    By choosing different resistances $K_{ij}$ between different nodes one
    can generate tighter or looser constraints on the average 
    distances between the points $i$ and $j$.
    }
    \label{fig:expanded}
\end{figure}
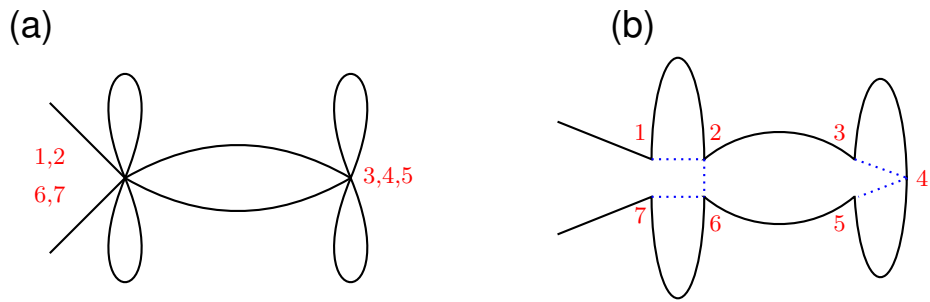

So far, we considered network topologies with constraints imposed by Dirac delta functions, 
see Eq.~\eqref{generaleqweight}. We denote these networks as ``tight''. This condition 
can be relaxed by adding a harmonic potential between some nodes, say $i$ and $j$. This amounts to using a weight 
\begin{equation}
    p_s = e^{-\frac{3(\vec{x}_i - \vec{x}_j)^2}{2K\sigma}}
\label{def:ps}
\end{equation}
instead of the $\delta$ functions. Here $K$ can be interpreted as an inverse spring
stiffness. As $K \to 0$ the nodes $i$ and $j$ are forced to get closer and closer,
recovering the tight network configuration in this limit.
Fig.~\ref{fig:expanded} shows a tight (a) and the ``expanded'' 
counterpart (b) of the same network topology. 
The heterogeneous inverse stiffness $K_{ij}$ can be chosen as a function of the node types $i$ and $j$, thus tuning the average distances between the
transcription units in the chromatin fibre. The analogy with Kirchhoff resistor networks can be applied to the extended network as well, by keeping the additional
resistances $K_{ij}$ into account.

\section{Combinatorial multiplicity of unlabelled loop networks}

In this Section, we will discuss how to count the number of labelled loop networks corresponding to a given unlabelled network. We refer to this number as the ``combinatorial multiplicity'' of the (unlabelled) network. We will consider here a general chain-like configuration (where clusters are joined up linearly), whereas Appendix A deals with the case of three-cluster configurations [which can either be chain-like or triangular, as in Fig.~\ref{fig2}(d)].

\begin{figure}[!h]
\begin{center}
\begin{tikzpicture}[node distance=2.cm,auto,main node/.style={fill,circle,draw,inner sep=0pt,minimum size=5pt}]
\node (1) {};
\node[main node] (2) [right of=1,xshift=-0.5cm] {};
\node[main node] (3) [right of=2] {};
\node (4) [right of=3,xshift=-0.5cm] {};

\node (5) [right of=4,xshift=-0.5cm] {};
\node[main node] (6) [right of=5,xshift=-10pt] {};
\node[main node] (7) [right of=6] {};
\node (8) [right of=7,xshift=-0.5cm] {};
\node (9) [below of=6,yshift=1.4cm] {{\fontsize{12}{0pt}$a$}};
\node (10) [below of=7,yshift=1.4cm] {{\fontsize{12}{0pt}$b$}};

\node (11) [right of=8,xshift=-1.0cm] {};
\node[main node] (12) [right of=11,xshift=-1.0cm] {};
\node[main node] (13) [right of=12] {};
\node (14) [right of=13,xshift=-0.5cm] {};
\node (15) [below of=12,yshift=1.4cm,xshift=-4pt] {{\fontsize{12}{0pt} $a$}};
\node (16) [below of=13,yshift=1.4cm,xshift=3pt] {{\fontsize{12}{0pt} $b$}};

\node (note) at (0.15,2.0) {{\fontsize{12}{0pt}\usefont{T1}{phv}{m}{n} (a)}};
\node (note) at (6.75,2.0) {{\fontsize{12}{0pt}\usefont{T1}{phv}{m}{n} (b)}};
\node (note) at (12.50,2.0) {{\fontsize{12}{0pt}\usefont{T1}{phv}{m}{n} (c)}};

\path
(1) edge (2)
(2) edge (3)
(3) edge (4)
(2) [bend right=40] edge (3)
(2) [bend left=40] edge (3);

\loopNarrowN{3}
\loopNarrowNW{2}
\loopNarrowN{2}

\path
(5)  [->,>=stealth] edge (6)
(7) [->,>=stealth] edge (8);

\path
(6) edge (7)
(6) [bend right=40] edge (7)
(6) [bend left=40] edge (7);

\loopNarrowN{7}
\loopNarrowNW{6}
\loopNarrowN{6}

\path
(13)  [->,>=stealth, bend left=80] edge (12);

\path
(13)  [->,>=stealth] edge (12)
(12)  [->,>=stealth, bend right=40] edge (13)
(12)  [->,>=stealth, bend left=40] edge (13);

\loopNarrowN{13}
\loopNarrowNW{12}
\loopNarrowN{12}

\end{tikzpicture}
\end{center}
\caption{Unlabelled networks. (a) Example of an unlabelled network configuration. (b) After directing the entry and exit edges, we would like to compute the number of possible traversals of the graphs, which equals the number of labelled networks corresponding to this unlabelled configuration. (c) Sketch of the gluing and ordering procedure, which we use to convert the unlabelled network into an Eulerian digraph.}
\label{fig:unlabellednetworks}
\end{figure}
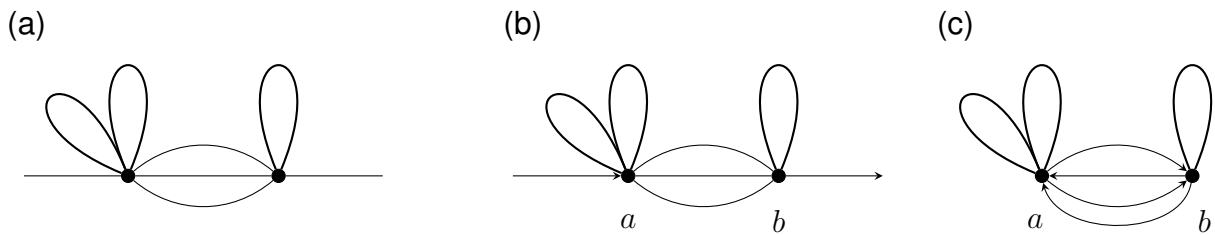

More specifically, we will assume in what follows that we are given a general configuration that is connected, and with clusters (or nodes) of even degree (except for the endpoints, which have degree 1), such as the one in Fig.~\ref{fig:unlabellednetworks}(a). The number of possible labellings is equal to the number of different traversals of the network given by the configuration under investigation. Note that, given our hypothesis that nodes are connected and have even degree, there is always at least one such traversal (this is a celebrated result in graph theory). A traversal is associated with a unique labelling of edges (and hence of nodes), as edges can be labelled following the order in which they are traversed. 

The underlying idea is that to find the combinatorial multiplicity of any given chromatin loop network, we need to map our configuration to a directed Eulerian graph. To do so, we first label the clusters, $a$ and $b$ in Fig.~\ref{fig:unlabellednetworks}(b). We then identify (glue) the entry and exit edges to obtain an {\em Eulerian graph}, that is a connected graph with each node of even degree. Our goal is then equivalent to computing the number of traversals (at least one) starting from the edge $b\rightarrow a$. For our running example, we look at the graph in Fig.~\ref{fig:unlabellednetworks}(b).

Counting traversals in undirected Eulerian graphs is a difficult problem. However, 
there is an elegant solution to counting traversals in {\em directed} Eulerian graphs (or {\em Eulerian digraphs}) via the ``BEST Theorem'' (see~\ref{BEST-sec}). 
A directed Eulerian graph is {\em strongly connected} if it is possible to get from any node to any other node following the edge directions, and it is {\em balanced} if each node has the same degree in and out. So, our methodology will be as follows. 
\begin{itemize}
\item[(1)] turn a given configuration into an Eulerian graph ${\mathcal G}$ in the way described above;
\item[(2)] consider all non-equivalent (up to permutations of multiple edges and loops, and possibly other symmetries) ways of orienting edges of ${\mathcal G}$ that results in an Eulerian digraph;
\item[(3)] apply the BEST Theorem in each of the possible orientation cases, and then sum up over all cases. 
\end{itemize}

In our calculation, we will take into account the fact that multiple edges, as well as multiple loops (i.e., edges with the same endpoints), are considered to be {\em indistinguishable}, meaning that the order in which we traverse multiple edges from a vertex $a$ to a vertex $b$ (or from $b$ to $a$) is ignored. 
In addition, we assume that the entrance to the network is fixed with a cluster $a$ to be visited first, and the exit of the network is fixed with a cluster $b$ to be visited last. For the above configuration, we have the following situation.

\subsection{The BEST Theorem}\label{BEST-sec}
The name ``BEST'' in the ``BEST Theorem'' is an acronym of the names of people who discovered it: N.\ G.\ de Bruijn, Tatyana Ehrenfest, Cedric Smith and W.\ T.\ Tutte.
The BEST Theorem \cite[Theorem 5b]{vA-deB1951}  (also see \cite{Stan1999} and \cite{Fred1982}) states that the number of Eulerian cycles in (traversals of) an Eulerian digraph with the vertex set $V$ and initial edge $e=v\rightarrow u$ is given by the formula
\begin{equation}\label{BEST-formula}
(\mbox{\# spanning trees rooted at }v)\cdot\prod_{x\in V}(\mbox{outdegree}(x)-1)!.
\end{equation}
Here we assume the spanning trees to be directed towards $b$, and it is known that the choice of $b$ is not important (the result will always be the same). 

An application of the BEST Theorem can be demonstrated on the configuration in Fig.~\ref{fig:unlabellednetworks} considered above. There is a unique (up to permutation of multiple edges) good orientation of the graph, shown in Fig.~\ref{fig:unlabellednetworks}(c).
There is also a single spanning tree (from $a$ to $b$), while  outdegree$(a)=4$, and  outdegree$(b)=3$. Hence, by the BEST Theorem [Eq.~(\ref{BEST-formula})], there are 
$$1\cdot 3!\cdot 2! = 12$$
Eulerian cycles (i.e.\ traversals) beginning with $b\rightarrow a$. However, recall that multiple edges going in the same direction are indistinguishable for us, and therefore, considering the presence of two multiple edges $a\rightarrow b$, the number of Eulerian cycles of interest is
$$\frac{12}{2!}=6.$$
The six traversals are:
$$
aaababb \qquad \, aaabbab \qquad \, aabbaab \qquad \, aabaabb \qquad \, abaaabb \qquad \, abbaaab \,. 
$$
Since the orientation of the configuration in Fig.~\ref{fig:unlabellednetworks} is unique, the combinatorial multiplicity of the corresponding unlabelled network is $6$.

\begin{figure}

\begin{center}

\begin{tikzpicture}[node distance=2.cm,auto,main node/.style={fill,circle,draw,inner sep=0pt,minimum size=3pt}]

\node (1) [yshift=-10pt] {};
\node[main node] (2) [right of=1,xshift=-5pt] {};
\node[main node] (x)  [right of=2] {};
\node[main node] (3) [right of=x] {};
\node (4) [right of=3,xshift=-5pt] {};
\node (5) [below of=2,yshift=1.4cm,xshift=-4pt] {{\fontsize{12}{0pt} $a$}};
\node (6) [below of=3,yshift=1.4cm,xshift=3pt] {{\fontsize{12}{0pt} $b$}};

\path
(1) edge (2)
(3) edge (4);

\path
(2) edge (x)
(2) [bend right=40] edge (x)
(2) [bend left=40] edge (x);

\path
(x) edge (3)
(x) [bend right=40] edge (3)
(x) [bend left=40] edge (3);

\loopNarrowN{x}
\loopNarrowN{2}




\node (11) [right of=4,xshift=-1.0cm] {};
\node[main node] (12) [right of=11,xshift=-5pt] {};
\node[main node] (1x)  [right of=12] {};
\node[main node] (13) [right of=1x] {};
\node (14) [right of=13,xshift=-5pt] {};
\node (15) [below of=12,yshift=1.4cm,xshift=-4pt] {{\fontsize{12}{0pt} $a$}};
\node (16) [below of=13,yshift=1.4cm,xshift=3pt] {{\fontsize{12}{0pt} $b$}};
\node (17) [below of=1x,yshift=1.4cm] {{\fontsize{12}{0pt} $c$}};

\path
(13)  [->,>=stealth, bend left=70] edge (12);

\path
(13)  [->,>=stealth] edge (1x)
(1x)  [->,>=stealth] edge (12);

\path
(12) [->,>=stealth, bend right=40] edge (1x)
(12) [->,>=stealth, bend left=40] edge (1x);

\path
(1x) [->,>=stealth, bend right=40] edge (13)
(1x) [->,>=stealth, bend left=40] edge (13);

\loopNarrowN{1x}
\loopNarrowN{12}

\node (note) at (0.15,2.0) {{\fontsize{12}{0pt}\usefont{T1}{phv}{m}{n} (a)}};
\node (note) at (10.40,2.0) {{\fontsize{12}{0pt}\usefont{T1}{phv}{m}{n} (b)}};

\end{tikzpicture}

\end{center}
\caption{A configuration (a) and the Eulerian digraph corresponding to it (b).}\label{conf-digraph-ex}
\end{figure}
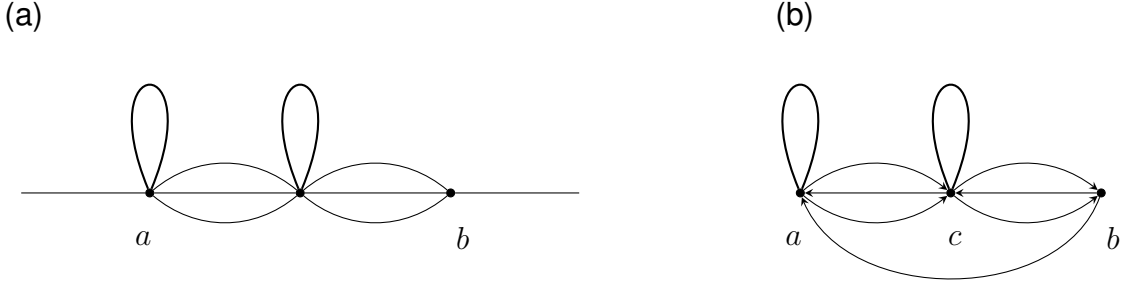

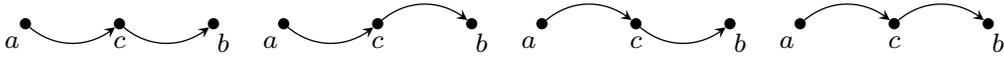
\begin{figure}

\begin{center}
\scalebox{1.25}{
\begin{tabular}{cccc}

\begin{tikzpicture}[node distance=1cm,auto,main node/.style={fill,circle,draw,inner sep=0pt,minimum size=3pt}]
\node[main node] (2) {};
\node[main node] (x)  [right of=2] {};
\node[main node] (3) [right of=x] {};
\node (5) [below of=2,yshift=0.8cm,xshift=-4pt] {{\footnotesize $a$}};
\node (6) [below of=3,yshift=0.8cm,xshift=3pt] {{\footnotesize $b$}};
\node (7) [below of=x,yshift=0.8cm] {{\footnotesize $c$}};

\path
(2) [->,>=stealth, bend right=40] edge (x);

\path
(x) [->,>=stealth, bend right=40] edge (3);
\end{tikzpicture}

&

\begin{tikzpicture}[node distance=1cm,auto,main node/.style={fill,circle,draw,inner sep=0pt,minimum size=3pt}]
\node[main node] (2) {};
\node[main node] (x)  [right of=2] {};
\node[main node] (3) [right of=x] {};
\node (5) [below of=2,yshift=0.8cm,xshift=-4pt] {{\footnotesize $a$}};
\node (6) [below of=3,yshift=0.8cm,xshift=3pt] {{\footnotesize $b$}};
\node (7) [below of=x,yshift=0.8cm] {{\footnotesize $c$}};

\path
(2) [->,>=stealth, bend right=40] edge (x);

\path
(x) [->,>=stealth, bend left=40] edge (3);
\end{tikzpicture}

&

\begin{tikzpicture}[node distance=1cm,auto,main node/.style={fill,circle,draw,inner sep=0pt,minimum size=3pt}]
\node[main node] (2) {};
\node[main node] (x)  [right of=2] {};
\node[main node] (3) [right of=x] {};
\node (5) [below of=2,yshift=0.8cm,xshift=-4pt] {{\footnotesize $a$}};
\node (6) [below of=3,yshift=0.8cm,xshift=3pt] {{\footnotesize $b$}};
\node (7) [below of=x,yshift=0.8cm] {{\footnotesize $c$}};

\path
(2) [->,>=stealth, bend left=40] edge (x);

\path
(x) [->,>=stealth, bend right=40] edge (3);
\end{tikzpicture}

&

\begin{tikzpicture}[node distance=1cm,auto,main node/.style={fill,circle,draw,inner sep=0pt,minimum size=3pt}]
\node[main node] (2) {};
\node[main node] (x)  [right of=2] {};
\node[main node] (3) [right of=x] {};
\node (5) [below of=2,yshift=0.8cm,xshift=-4pt] {{\footnotesize $a$}};
\node (6) [below of=3,yshift=0.8cm,xshift=3pt] {{\footnotesize $b$}};
\node (7) [below of=x,yshift=0.8cm] {{\footnotesize $c$}};

\path
(2) [->,>=stealth, bend left=40] edge (x);

\path
(x) [->,>=stealth, bend left=40] edge (3);
\end{tikzpicture}

\end{tabular}
}
\end{center}
\caption{The spanning trees rooted at node $a$ in Fig.~\ref{conf-digraph-ex} to the right.}\label{spanning-trees-for-ex}
\end{figure}

As another application, we consider Fig.~\ref{conf-digraph-ex}. There are four spanning trees rooted at node $a$ in Fig.~\ref{conf-digraph-ex} to the right that are presented in Fig.~\ref{spanning-trees-for-ex}. Moreover, for that graph outdegree$(a)=3$, outdegree$(c)=4$ and outdegree$(b)=2$. Hence, by  the BEST Theorem (formula (\ref{BEST-formula})), there are 
$$4\cdot 2!\cdot 3!\cdot 1! = 48$$
Eulerian cycles (i.e.\ traversals) beginning with $v\rightarrow u$. Accounting for the fact that multiple edges going in the same direction are indistinguishable, and considering the presence of two multiple edges $a\rightarrow c$ and two multiple edges $c\rightarrow b$, the number of Eulerian cycles of interest is
$$\frac{48}{2!\cdot 2!}=12.$$
Since the orientation of the configuration in Fig.~\ref{conf-digraph-ex} is unique, $12$ is the final combinatorial multiplicity for the network. 

\subsection{Chain-like configurations}

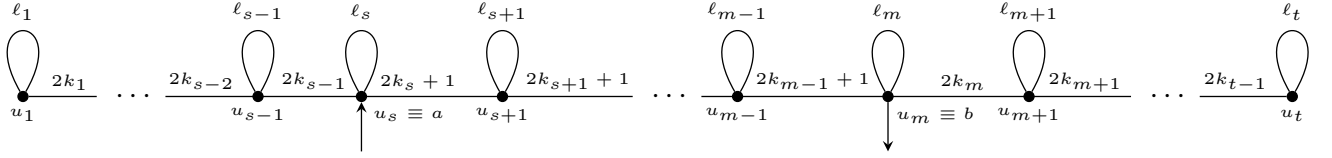
\begin{figure}

\begin{center}

\scalebox{1.25}{
\begin{tikzpicture}[node distance=1cm,auto,main node/.style={fill,circle,draw,inner sep=0pt,minimum size=3pt}]

\node[main node] (1) {};
\node (2) [right of=1,xshift=-0.1cm] {};
\node (3) [right of=2,xshift=-0.7cm] {$\cdots$};
\node (4) [right of=3,xshift=-0.8cm] {};
\node (5)[main node]  [right of=4,xshift=0.1cm] {};
\node (6)[main node]  [right of=5,xshift=0.1cm] {};
\node (7)[main node]  [right of=6,xshift=0.5cm] {};
\node (8)  [right of=7,xshift=0.5cm] {};
\node (9) [right of=8,xshift=-0.7cm] {$\cdots$};
\node (10) [right of=9,xshift=-0.8cm] {};
\node (11)[main node]  [right of=10,xshift=-0.5cm] {};
\node (12)[main node]  [right of=11,xshift=0.6cm] {};
\node (13)[main node]  [right of=12,xshift=0.5cm] {};
\node (14)  [right of=13,xshift=0.2cm] {};
\node (15) [right of=14,xshift=-0.7cm] {$\cdots$};
\node (16) [right of=15,xshift=-0.8cm] {};
\node (17)[main node]  [right of=16,xshift=0.1cm] {};

\node (18)  [below of=6,yshift=0.3cm] {};
\node (19)  [below of=12,yshift=0.3cm] {};

\node [above of=1, yshift=-0.1cm] {{\tiny $\ell_1$}};
\node [above of=5, yshift=-0.1cm] {{\tiny $\ell_{s-1}$}};
\node [above of=6, yshift=-0.1cm] {{\tiny $\ell_s$}};
\node [above of=7, yshift=-0.1cm] {{\tiny $\ell_{s+1}$}};
\node [above of=11, yshift=-0.1cm] {{\tiny $\ell_{m-1}$}};
\node [above of=12, yshift=-0.1cm] {{\tiny $\ell_m$}};
\node [above of=13, yshift=-0.1cm] {{\tiny $\ell_{m+1}$}};
\node [above of=17, yshift=-0.1cm] {{\tiny $\ell_t$}};

\node [below of=1, yshift=0.8cm] {{\tiny $u_1$}};
\node [below of=5, yshift=0.8cm] {{\tiny $u_{s-1}$}};
\node [below of=6, yshift=0.8cm,xshift=0.5cm] {{\tiny $u_s \equiv a$}};
\node [below of=7, yshift=0.8cm] {{\tiny $u_{s+1}$}};
\node [below of=11, yshift=0.8cm] {{\tiny $u_{m-1}$}};
\node [below of=12, yshift=0.8cm,xshift=0.5cm] {{\tiny $u_m\equiv b$}};
\node [below of=13, yshift=0.8cm] {{\tiny $u_{m+1}$}};
\node [below of=17, yshift=0.8cm] {{\tiny $u_t$}};

\node [above of=1, yshift=-0.85cm,xshift=0.5cm] {{\tiny $2k_1$}};
\node [above of=4, yshift=-0.85cm,xshift=0.5cm] {{\tiny $2k_{s-2}$}};
\node [above of=5, yshift=-0.85cm,xshift=0.6cm] {{\tiny $2k_{s-1}$}};
\node [above of=6, yshift=-0.85cm,xshift=0.6cm] {{\tiny $2k_s+1$}};
\node [above of=7, yshift=-0.85cm,xshift=0.8cm] {{\tiny $2k_{s+1}+1$}};
\node [above of=11, yshift=-0.85cm,xshift=0.8cm] {{\tiny $2k_{m-1}+1$}};
\node [above of=12, yshift=-0.85cm,xshift=0.8cm] {{\tiny $2k_m$}};
\node [above of=13, yshift=-0.85cm,xshift=0.6cm] {{\tiny $2k_{m+1}$}};
\node [above of=16, yshift=-0.85cm,xshift=0.5cm] {{\tiny $2k_{t-1}$}};

\path
(18) [->,>=stealth] edge (6)
(12) [->,>=stealth] edge (19);

\path
(1) edge (2)
(4) edge (8)
(10) edge (14)
(16) edge (17);

\path
(1) [loop above] edge [in=120,out=60,loop] (1)
(5) [loop above] edge [in=120,out=60,loop] (5)
(6) [loop above] edge [in=120,out=60,loop] (6)
(7) [loop above] edge [in=120,out=60,loop] (7)
(11) [loop above] edge [in=120,out=60,loop] (11)
(12) [loop above] edge [in=120,out=60,loop] (12)
(13) [loop above] edge [in=120,out=60,loop] (13)
(17) [loop above] edge [in=120,out=60,loop] (17);
\end{tikzpicture}
}

\end{center}
\caption{The most general chain configuration, where the labels of edges ($2k_i(+1)$) and loops  ($\ell_j$) give their multiplicities and the clusters are labelled by $\{u_1,\ldots,u_t\}$; we assume, without loss of generality, that $1\leq s\leq m\leq t$.}\label{most-general-path-like}
\end{figure}

A useful case to consider is that of chain-like configurations, where clusters are connected linearly. In other words, for a generic number of clusters $t$, if we label the clusters $\{u_1,\ldots,u_t\}$, an internal cluster  $u_i$ ($i=2,\ldots,t-1$) is only connected to $u_{i\pm 1}$ (while $u_1$ is connected only to $u_2$ and $u_t$ only to $u_{t-1}$). The most general chain-like configuration is given in Fig.~\ref{most-general-path-like}, 
where for simplicity of notation the multiplicities of loops are denoted by $\ell_i$ for a cluster $u_i$, and $2k_i$ and $2k_i+1$, $k_i\geq 1$, denote the multiplicities of edges between the clusters $u_i$ and $u_{i+1}$. We note that the multiplicities of edges are forced to be either even or odd in particular parts of the configuration to ensure that at least one traversal of the network exists.  Also, $u_s=a$ is the entrance cluster, and we exit from cluster $u_m=b$, where, without loss of generality (by symmetry), we assume that $1\leq s\leq m\leq t$.

We next use the BEST Theorem to show that the combinatorial multiplicity (number of traversals) of the configuration in Fig.~\ref{most-general-path-like} is
\begin{equation}\label{most-general-case-formula}
\frac{\displaystyle\prod_{i=1}^{s-1}k_i\displaystyle\prod_{j=s}^{m-1}(k_j+1)\displaystyle\prod_{x=m}^{t-1}k_x\displaystyle\prod_{r=1}^{s-1}(k_{r-1}+k_r+\ell_r-1)!\displaystyle\prod_{z=s}^{m}(k_{z-1}+k_z+\ell_z)!\displaystyle\prod_{p=m+1}^{t}(k_{p-1}+k_p+\ell_p-1)!}{\displaystyle\prod_{i=1}^{t}\ell_i!\displaystyle\prod_{j=1}^{s-1}(k_j!)^2\displaystyle\prod_{r=s}^{m-1}k_r!(k_r+1)!\displaystyle\prod_{p=m}^{t-1}(k_p!)^2}
\end{equation}
where we define $k_0=0$ and $k_t=0$. 

We can see that a possible orientation of the configuration in Figure~\ref{most-general-path-like} is unique (up to a permutation of multiple edges), while the number of spanning trees rooted in $u_m=b$ equals
$$\displaystyle\prod_{i=1}^{s-1}k_i\displaystyle\prod_{j=s}^{m-1}(k_j+1)\displaystyle\prod_{x=m}^{t-1}k_x.$$
We can now use (\ref{BEST-formula}) to prove Eq.~(\ref{most-general-case-formula}) after considering multiplicities and observing that
\begin{small}
\begin{eqnarray}
\mbox{outdegree}(u_1) &=& k_1+\ell_1 \nonumber\\
\mbox{outdegree}(u_2) &=& k_1+k_2 + \ell_2 \nonumber\\
&\vdots & \nonumber \\
\mbox{outdegree}(u_{s-1}) &=& k_{s-2}+k_{s-1} + \ell_{s-1} \nonumber\\
\mbox{outdegree}(u_{s}) &=& k_{s-1}+k_{s} + \ell_{s} +1 \nonumber\\
\mbox{outdegree}(u_{s+1}) &=& k_{s}+k_{s+1} + \ell_{s+1}+1 \nonumber\\
&\vdots & \nonumber \\
\mbox{outdegree}(u_{m-1}) &=& k_{m-2}+k_{m-1} + \ell_{m-1}+1 \nonumber\\
\mbox{outdegree}(u_{m}) &=& k_{m-1}+k_{m} + \ell_{m}+1 \mbox{ \ \ (``+1'' here is the exit edge)}\nonumber\\
\mbox{outdegree}(u_{m+1}) &=& k_{m}+k_{m+1} + \ell_{m+1} \nonumber\\
\mbox{outdegree}(u_{m+2}) &=& k_{m+1}+k_{m+2} + \ell_{m+2} \nonumber\\
&\vdots & \nonumber \\
\mbox{outdegree}(u_{t-1}) &=& k_{t-2}+k_{t-1} + \ell_{t-1} \nonumber\\
\mbox{outdegree}(u_t) &=& k_{t-1}+\ell_t \nonumber
\end{eqnarray}
\end{small}
Note that in the denominator in (\ref{most-general-case-formula}) the factor of $(k_j!)^2$ (resp., $(k_p!)^2$) comes from the fact that the number of multiple edges $u_j\rightarrow u_{j+1}$ (resp., $u_p\rightarrow u_{p+1}$) is the same as that of multiple edges $u_{j+1}\rightarrow u_{j}$ (resp., $u_{p+1}\rightarrow u_{p}$).

Next, we discuss particular cases of Eq.~(\ref{most-general-case-formula}) when the number of clusters is 2 or 3 (one cluster is trivial). 

\subsubsection{Chain-like configurations with two clusters}

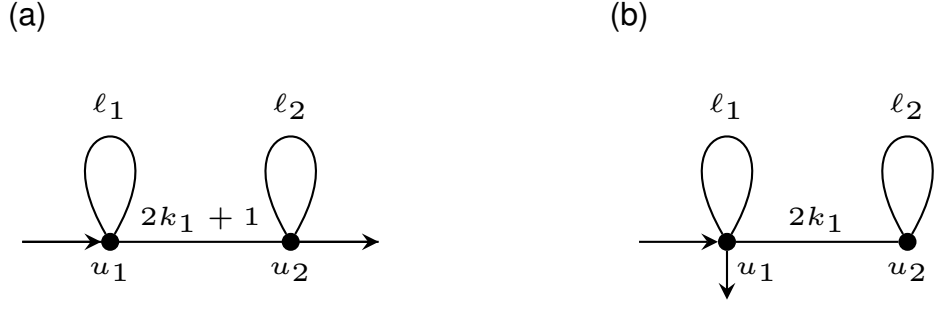
\begin{figure}
\begin{center}

\scalebox{2.0}{
\begin{tikzpicture}[node distance=1cm,auto,main node/.style={fill,circle,draw,inner sep=0pt,minimum size=3pt}]

\node (1) {};
\node (2)[main node] [right of=1,xshift=-0.3cm] {};
\node (4)[main node] [right of=2,xshift=0.2cm] {};
\node (5)  [right of=4,xshift=-0.3cm] {};
\node (6)  [below of=2,yshift=0.8cm] {{\tiny $u_1$}};
\node (7)  [below of=4,yshift=0.8cm] {{\tiny $u_2$}};

\node (note) at (0.15,1.5) {{\fontsize{6}{0pt}\usefont{T1}{phv}{m}{n} (a)}};

\path
(1) edge (5);

\path
(2) [loop above] edge [in=120,out=60,loop] (2)
(4) [loop above] edge [in=120,out=60,loop] (4);

\node [above of=2, yshift=-0.1cm] {{\tiny $\ell_1$}};
\node [above of=4, yshift=-0.1cm] {{\tiny $\ell_2$}};

\node [above of=2, yshift=-0.85cm,xshift=0.6cm] {{\tiny $2k_1+1$}};

\path
(1) [->,>=stealth] edge (2)
(4) [->,>=stealth] edge (5);


\node (8) [right of=5, xshift=0.5cm] {};
\node (9)[main node] [right of=8,xshift=-0.3cm] {};
\node (10)[main node] [right of=9,xshift=0.2cm] {};
\node (11)  [below of=9,yshift=0.5cm] {};
\node (12)  [below of=9,yshift=0.8cm,xshift=0.2cm] {{\tiny $u_1$}};
\node (13)  [below of=10,yshift=0.8cm] {{\tiny $u_2$}};

\node (note) at (4.15,1.5) {{\fontsize{6}{0pt}\usefont{T1}{phv}{m}{n} (b)}};

\path
(9) edge (10);

\path
(9) [loop above] edge [in=120,out=60,loop] (9)
(10) [loop above] edge [in=120,out=60,loop] (10);

\node [above of=9, yshift=-0.1cm] {{\tiny $\ell_1$}};
\node [above of=10, yshift=-0.1cm] {{\tiny $\ell_2$}};

\node [above of=9, yshift=-0.85cm,xshift=0.6cm] {{\tiny $2k_1$}};

\path
(8) [->,>=stealth] edge (9)
(9) [->,>=stealth] edge (11);

 \end{tikzpicture} 
}
 
\end{center}
\caption{Types of inequivalent $2$-cluster chain-like configurations discussed in the text.}
\label{chain-2-clusters}
\end{figure}

In the case of two clusters, there are only two possibilities up to symmetries, shown in Fig.~\ref{chain-2-clusters}. 

For configurations as in Fig.~\ref{chain-2-clusters} in which case Eq.~(\ref{most-general-case-formula}) simplifies as
 \begin{equation}\label{2-chain-1}\frac{(k_1+\ell_1)!(k_1+\ell_2)!}{\ell_1!\ell_2!(k_1!)^2}\end{equation}
 for the number of traversals, whereas for configurations as in Fig.~\ref{chain-2-clusters}, the combinatorial multiplicity is
 \begin{equation}\label{2-chain-2}\frac{(k_1+\ell_1)!(k_1+\ell_2-1)!}{k_1\ell_1!\ell_2!((k_1-1)!)^2}.\end{equation}

 For example, in the case of $n=8$ TUs (considered in \cite{short,Bonato2024}), $k_1=1$, $\ell_1=\ell_2=2$,  the formula (\ref{2-chain-1}) gives nine traversals, which are, for $u_1=a$ and $u_2=b$,
\begin{eqnarray}\nonumber
aaababbb \qquad aabaabbb \qquad  abaaabbb \\
aaabbabb \qquad aabbaabb \qquad abbaaabb  \\
aaabbbab \qquad aabbbaab \qquad  abbbaaab
\end{eqnarray}
while in the case of $n=8$, $k_1=1$, $\ell_1=2$ and $\ell_2=3$, Eq.~(\ref{2-chain-2}) gives three traversal, which are, for $u=a$ and $u_2=b$ 
\begin{equation}
aaabbbba \qquad aabbbbaa \qquad abbbbaaa \, .
\end{equation}

\subsubsection{Chain-like configurations with three clusters}\label{chain-conf-3-clust-sec}

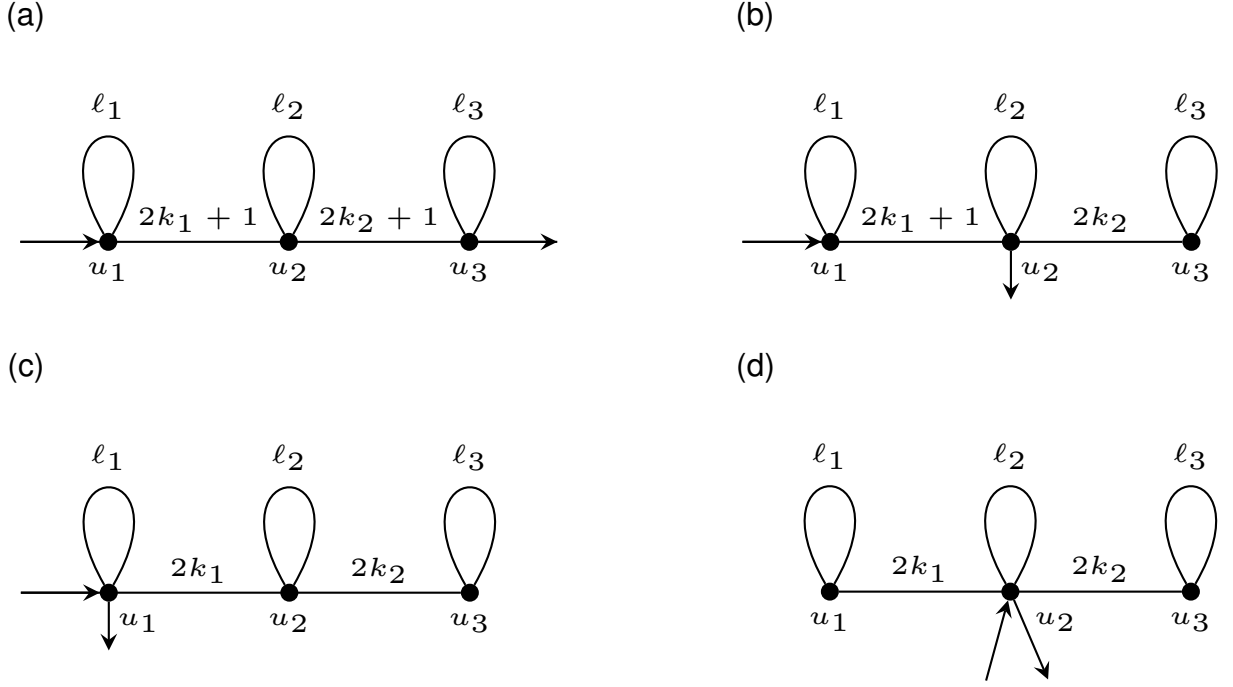
\begin{figure}
\begin{center}
\scalebox{2.0}{
\begin{tikzpicture}[node distance=1cm,auto,main node/.style={fill,circle,draw,inner sep=0pt,minimum size=3pt}]

\node (1) {};
\node (2)[main node] [right of=1,xshift=-0.3cm] {};
\node (3)[main node] [right of=2,xshift=0.2cm] {};
\node (4)[main node] [right of=3,xshift=0.2cm] {};
\node (5)  [right of=4,xshift=-0.3cm] {};
\node (6)  [below of=2,yshift=0.8cm] {{\tiny $u_1$}};
\node (7)  [below of=3,yshift=0.8cm] {{\tiny $u_2$}};
\node (8)  [below of=4,yshift=0.8cm] {{\tiny $u_3$}};

\node (note) at (0.15,1.5) {{\fontsize{6}{0pt}\usefont{T1}{phv}{m}{n} (a)}};

\path
(1) edge (5);

\path
(2) [loop above] edge [in=120,out=60,loop] (2)
(3) [loop above] edge [in=120,out=60,loop] (3)
(4) [loop above] edge [in=120,out=60,loop] (4);

\node [above of=2, yshift=-0.1cm] {{\tiny $\ell_1$}};
\node [above of=3, yshift=-0.1cm] {{\tiny $\ell_2$}};
\node [above of=4, yshift=-0.1cm] {{\tiny $\ell_3$}};

\node [above of=2, yshift=-0.85cm,xshift=0.6cm] {{\tiny $2k_1+1$}};
\node [above of=3, yshift=-0.85cm,xshift=0.6cm] {{\tiny $2k_2+1$}};

\path
(1) [->,>=stealth] edge (2)
(4) [->,>=stealth] edge (5);

\node (9) [right of=5, xshift=0.cm] {};
\node (10)[main node] [right of=9,xshift=-0.3cm] {};
\node (11)[main node] [right of=10,xshift=0.2cm] {};
\node (12)[main node] [right of=11,xshift=0.2cm] {};
\node (13)  [below of=11,yshift=0.5cm] {};
\node (14)  [below of=10,yshift=0.8cm] {{\tiny $u_1$}};
\node (15)  [below of=11,yshift=0.8cm,xshift=0.2cm] {{\tiny $u_2$}};
\node (16)  [below of=12,yshift=0.8cm] {{\tiny $u_3$}};

\node (note) at (5.0,1.5) {{\fontsize{6}{0pt}\usefont{T1}{phv}{m}{n} (b)}};

\path
(9) edge (12);

\path
(10) [loop above] edge [in=120,out=60,loop] (10)
(11) [loop above] edge [in=120,out=60,loop] (11)
(12) [loop above] edge [in=120,out=60,loop] (12);

\node [above of=10, yshift=-0.1cm] {{\tiny $\ell_1$}};
\node [above of=11, yshift=-0.1cm] {{\tiny $\ell_2$}};
\node [above of=12, yshift=-0.1cm] {{\tiny $\ell_3$}};

\node [above of=10, yshift=-0.85cm,xshift=0.6cm] {{\tiny $2k_1+1$}};
\node [above of=11, yshift=-0.85cm,xshift=0.6cm] {{\tiny $2k_2$}};

\path
(9) [->,>=stealth] edge (10)
(11) [->,>=stealth] edge (13);

 \end{tikzpicture} 
 }\\
\scalebox{2.0}{
\begin{tikzpicture}[node distance=1cm,auto,main node/.style={fill,circle,draw,inner sep=0pt,minimum size=3pt}]

\node (1) {};
\node (2)[main node] [right of=1,xshift=-0.3cm] {};
\node (3)[main node] [right of=2,xshift=0.2cm] {};
\node (4)[main node] [right of=3,xshift=0.2cm] {};
\node (5)  [below of=2,yshift=0.5cm] {};
\node (6)  [below of=2,yshift=0.8cm,xshift=0.2cm] {{\tiny $u_1$}};
\node (7)  [below of=3,yshift=0.8cm] {{\tiny $u_2$}};
\node (8)  [below of=4,yshift=0.8cm] {{\tiny $u_3$}};

\node (note) at (0.15,1.5) {{\fontsize{6}{0pt}\usefont{T1}{phv}{m}{n} (c)}};

\path
(1) edge (4);

\path
(2) [loop above] edge [in=120,out=60,loop] (2)
(3) [loop above] edge [in=120,out=60,loop] (3)
(4) [loop above] edge [in=120,out=60,loop] (4);

\node [above of=2, yshift=-0.1cm] {{\tiny $\ell_1$}};
\node [above of=3, yshift=-0.1cm] {{\tiny $\ell_2$}};
\node [above of=4, yshift=-0.1cm] {{\tiny $\ell_3$}};

\node [above of=2, yshift=-0.85cm,xshift=0.6cm] {{\tiny $2k_1$}};
\node [above of=3, yshift=-0.85cm,xshift=0.6cm] {{\tiny $2k_2$}};

\path
(1) [->,>=stealth] edge (2)
(2) [->,>=stealth] edge (5);

\node (9) [right of=4, xshift=2.4cm, yshift=-0.7cm]{};
\node (10)[main node] [above left of=9,xshift=-0.3cm] {};
\node (11)[main node] [right of=10,xshift=0.2cm] {};
\node (12)[main node] [right of=11,xshift=0.2cm] {};
\node (13)  [below of=11,yshift=0.3cm,xshift=0.3cm] {};
\node (14)  [below of=10,yshift=0.8cm] {{\tiny $u_1$}};
\node (15)  [below of=11,yshift=0.8cm,xshift=0.3cm] {{\tiny $u_2$}};
\node (16)  [below of=12,yshift=0.8cm] {{\tiny $u_3$}};

\path
(10) edge (12);

\path
(10) [loop above] edge [in=120,out=60,loop] (10)
(11) [loop above] edge [in=120,out=60,loop] (11)
(12) [loop above] edge [in=120,out=60,loop] (12);

\node (note) at (5.0,1.5) {{\fontsize{6}{0pt}\usefont{T1}{phv}{m}{n} (d)}};

\node [above of=10, yshift=-0.1cm] {{\tiny $\ell_1$}};
\node [above of=11, yshift=-0.1cm] {{\tiny $\ell_2$}};
\node [above of=12, yshift=-0.1cm] {{\tiny $\ell_3$}};

\node [above of=10, yshift=-0.85cm,xshift=0.6cm] {{\tiny $2k_1$}};
\node [above of=11, yshift=-0.85cm,xshift=0.6cm] {{\tiny $2k_2$}};

\path
(9) [->,>=stealth] edge (11)
(11) [->,>=stealth] edge (13);

 \end{tikzpicture} 
 }
 
 \end{center}
 \caption{Types of inequivalent 3-cluster chain-like configurations discussed in the text.}
 \label{chain-3-clusters}
 \end{figure}

In the case of three clusters, there are four possible types of configurations, shown in Fig.~\ref{chain-3-clusters}.
For the configuration
 
 For the case in Fig.~\ref{chain-3-clusters}(a), we can use Eq.~(\ref{most-general-case-formula}) with $s=1$, $m=3$, $t=m$, hence  the combinatorial multiplicity is
 \begin{equation}\label{3-chain-1} 
 \frac{(k_1+\ell_1)!(k_1+k_2+\ell_2)!(k_2+\ell_3)!}{\ell_1!\ell_2!\ell_3!(k_1!)^2(k_2!)^2}.
 \end{equation}
 
 
For configurations of the type in Fig.~\ref{chain-3-clusters}(b), Eq.~(\ref{most-general-case-formula}) simplifies to
 \begin{equation}\label{3-chain-2} 
  \frac{(k_1+\ell_1)!(k_1+k_2+\ell_2)!(k_2+\ell_3-1)!}{\ell_1!\ell_2!\ell_3!(k_1!)^2(k_2-1)!k_2!}
\end{equation}
since in this case $s=1$, $m=2$ and $t=3$. 

 
 For the case of Fig.~\ref{chain-3-clusters}(c), $s=m=1$ and $t=3$, and we obtain the following formula for the combinatorial multiplicity,
 \begin{equation}\label{3-chain-3} 
  \frac{(k_1+\ell_1)!(k_1+k_2+\ell_2-1)!(k_2+\ell_3-1)!}{\ell_1!\ell_2!\ell_3!(k_1-1)!k_1!(k_2-1)!k_2!} .
 \end{equation}

 
 Finally, for the configuration in Fig.~\ref{chain-3-clusters}(d), 
 Eq.~(\ref{most-general-case-formula}) simplifies as
 \begin{equation}\label{3-chain-4} 
 \frac{(k_1+\ell_1-1)!(k_1+k_2+\ell_2)!(k_2+\ell_3-1)!}{\ell_1!\ell_2!\ell_3!(k_1-1)!k_1!(k_2-1)!k_2!}
 \end{equation}
since $s=m=2$ and $t=3$ in this case. 

\section{Structural diversity of chromatin loop networks}

In this Section, we apply the theoretical frameworks developed in the previous Section to find the Shannon entropy, or structural diversity, of a set of polydisperse chromatin loop networks with a given set of distances between $n$ TUs, $\{l_1,\ldots,l_{n-1}\}$, which we refer to as a model ``gene topos''~\cite{Chiang2024}. We then explore how the structural diversity depends on the choice of $\{l_1,\ldots,l_{n-1}\}$, and how the measurement of structural diversity changes if we consider labelled or unlabelled networks. For concreteness, we focus on the case of $2$-cluster and $3$-cluster networks.

\subsection{Two-cluster configurations without singletons}
\begin{figure}[t]
    \centering
    \includegraphics[width=1\linewidth]{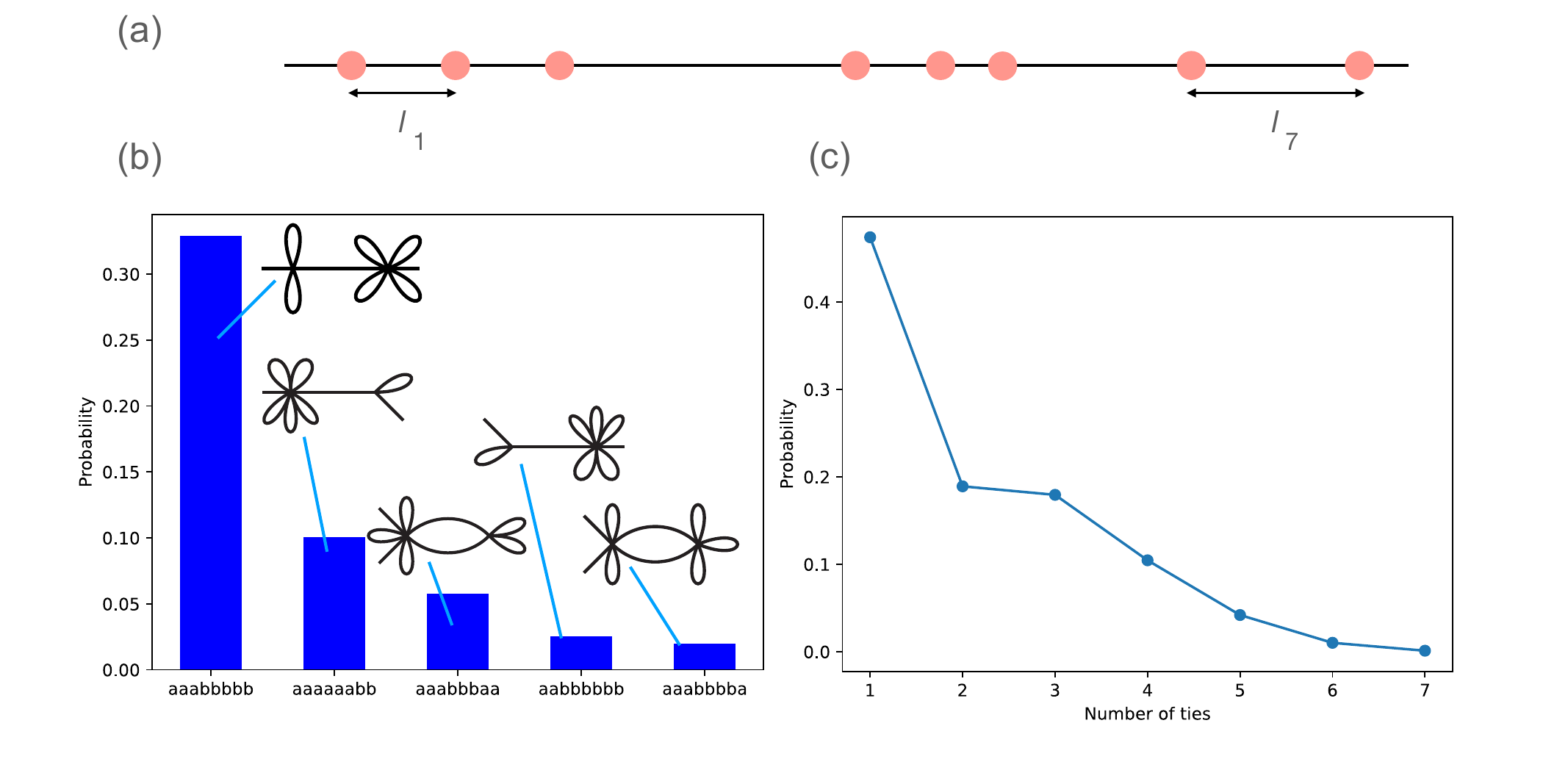}
    \caption{(a) We consider a chromatin fibre with $8$ transcription units, which have variable spacing between them. 
    The sketch is a generic example. The data shown in this figure are for the following choice of TUs spacing:
    $\{l_i\}_{i=1,\ldots,7}=\{6,6,33,4,2,15,6\}$, generated from a Poisson distribution with average 
    $\langle l \rangle =10$.
    (b) Curve showing the probability, or relative labelled topological weight with respect to the sum of 
    the weights, for the most likely topologies. Note that the fifth topology shown, $aaabbbba$, has the same probability as the $abbaaaaa$ topology. 
    (c) Probability of configurations with a given number of ties $n_t$. Each value of $n_t$ is associated with a single value of the rosette score ${\mathcal R}$: $n_t=1$ and $n_t=7$ respectively correspond to ${\mathcal R}=1$ and ${\mathcal R}=0$.}
    \label{fig:structuraldiversity01}
\end{figure}

We begin with the case of two-clusters labelled networks, without singletons. As discussed in~\cite{Bonato2024}, there are 
\begin{equation}
    N(n,2)=2^{n-1}-n-1
\end{equation}
two-cluster networks for a fibre with $n$ TUs. Note that we use the notation in~\cite{Bonato2024}, where we call $N(n,n_c)$ the number of networks with $n$ TUs and $n_c$ clusters. For our explicit calculation, we restrict ourselves to the case of $n=8$ [see Fig.~\ref{fig:structuraldiversity01}(a)], where all topologies can be exhaustively enumerated, and which is biologically relevant~\cite{short}.

Fig.~\ref{fig:structuraldiversity01}(b) shows the five labelled topologies with the highest topological weights,
or probability of occurrence, for a polydisperse network where the distance between neighbouring transcription
units is drawn from a Poisson distribution (with average equal to $10$ in the case shown with segment lengths
equal to $\{l_i\}_{i=1,\ldots,7}=\{6,6,33,4,2,15,6\}$). It can be seen that 
the most frequent topologies are rosette-like, as in the monodisperse case~\cite{Bonato2024}. Notably, the top
five topologies (out of $119$ possible) account for more than half of the total weight, so they are expected to appear in more than half of the configurations of the topos under consideration. The dominance of rosette structures is 
also apparent from an analysis of unlabelled loop networks, where we have coarse grained structures such that 
they are only distinguished by the number of ties between the clusters. Rosettes, which have $n_t=1$, have an
occurrence probability larger than $40\%$ for the distance distribution shown in Fig.~\ref{fig:structuraldiversity01},
and this is typical of a generic polydisperse network. 

One can understand some general features of the results of Fig.~\ref{fig:structuraldiversity01}(b) 
from a simple analysis. According to the definition \eqref{generaleqweight}, loops contribute
with a weight equal to one to $Z_{\cal G}$: therefore, the topological weight of a network only depends on the ties topology. For conformations with a single tie of length $l_i$, the 
topological weight is  $Z_{\cal G} \sim Z_i \sim l_i^{3/2}$, with $Z_i$ defined by
\eqref{def:Zi}. The highest weight corresponds to choosing the longest segment for the tie 
location. In our example $\max\{l_i\} =l_3 = 33$, 
which explains why the configuration with the topology with the highest weight is 
$aaabbbbb$: the first three TUs that are separated by segments of short length ($l_1=l_2=6$) are in the same $a$-cluster, and the 
remaining five TUs are in the same $b$-cluster with the two clusters being separated by the segment $l_3$.
The second highest probability for the single tie topology corresponds to fixing the tie at the second 
longest segment, which is $l_6=15$. This gives the topology $aaaaaabb$. The ratio of the probabilities of the two most probable topologies is
\begin{equation}
    \frac{Z_{aaabbbbb}}{Z_{aaaaaabb}} = \left( \frac{l_3}{l_6} \right)^{3/2} \simeq 3.26 \, .
\end{equation}
The third most probable topology with a single tie chooses the third longest segment for the tie location. We note that there are three segments of equal length $l_1=l_2=l_7=6$. However the only possible topology
without singletons (unbound TUs) is $aabbbbbb$, with the first two TUs in one cluster and the last six 
TUs in another cluster, as seen in Fig.~\ref{fig:structuraldiversity01}(b). 

For the topologies with two ties of length $l_i$ and $l_j$ the weight is obtained by placing the equivalent resistors in parallel, hence $Z_{\cal G} \sim [l_i l_j/(l_i+l_j)]^{3/2}$. The highest weight corresponds to
placing the ties where the segments have the highest lengths, which is $l_3=33$ and $l_6=16$ in our example. 
This is indeed the configuration $aaabbbaa$, which is the two-ties topology with the highest weight in 
Fig.~\ref{fig:structuraldiversity01}(b). The two-ties topology with the second highest weight is that with 
ties $l_3=33$ and $l_7=6$, which is $aaabbbba$. The ratio of their weights is
\begin{equation}
    \frac{Z_{aaabbbaa}}{Z_{aaabbbba}} = \left( \frac{l_3 l_6}{l_3+l_6} \, \frac{l_3 +l_7}{l_3 l_7}\right)^{3/2}
    \simeq 2.89 \, .
\end{equation}

We define the Shannon entropy or structural diversity of a set of $N$ labelled networks as follows,
\begin{equation}
    {\mathcal S}\equiv \frac{S}{k_b}=-\sum_{i=1}^{N}p_i \ln(p_i),
\end{equation}
where $p_i$ is the probability of occurrence of the $i$-th labelled topology, which equals its weight 
divided by the total weight -- i.e., the sum of all labelled topology weights.  
In our case, the sum goes up to $N=N(8,2)=119$.

It is useful to define the rosette score as~\cite{short,Brackley2016}
\begin{equation}
    {\mathcal R} = 1-\frac{n_t-n_c+1}{n-n_c} = \frac{n-(n_t+1)}{n-n_c},
\end{equation}
where $n_t$ is the total number of ties between the clusters. It can be seen that for a string of two rosettes, $n_t=1$, $n_c=2$ and ${\mathcal R}=1$. Instead, for a watermelon with $n=8$, the number of ties is $n_t=n-1=7$, so that $n=n_t+1$ and ${\mathcal R}=0$. 
More generally, the parameter ${\mathcal R}$ decreases with the number of ties, so it is minimal for a string of rosettes, and it measures in a continuous way the relative dominance of local over non-local loops.

\begin{figure}
    \centering
    \includegraphics[width=0.5\linewidth]{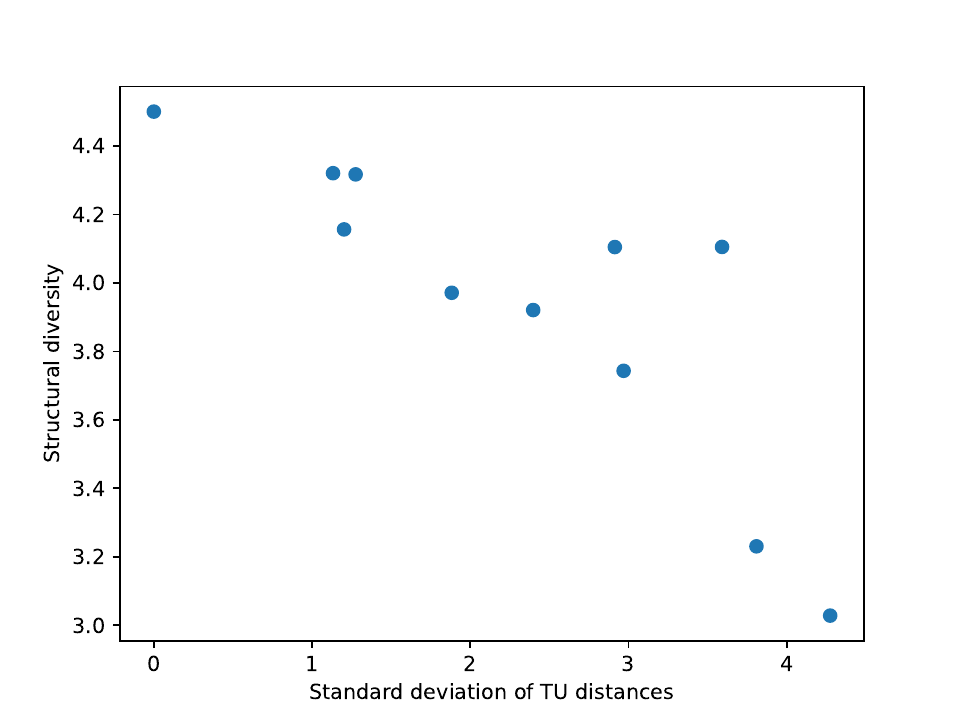}
    \caption{Scatter plot showing the structural diversity as a function of the standard deviation of the distance between neighbouring TUs, found for a number of different TU dispositions taken from a Poissonian with average distance $\bar l=10$.}
    \label{fig:structuraldiversity02}
\end{figure}

For the networks resulting from the 1D TU positioning in Fig.~\ref{fig:structuraldiversity01}, the structural 
diversity is ${\mathcal S}\simeq 3.38$, which is, as expected, lower than the Shannon entropy of a corresponding set of equally likely loop networks, $\log 119\simeq 4.78$. It is interesting to explore how the structural diversity changes for different patterns of 1D TU positioning. By analysing several TU positioning where the distance between successive TU is drawn from a Poisson distribution with the same average, we find that ${\mathcal S}$ decreases with the standard deviation of $\{l_i\}_{i=1,\ldots,8}$ (Fig.~\ref{fig:structuraldiversity02}): in other words, the largest structural diversity is obtained with uniform TU spacing. This is because if TUs have the same spacing, then more networks are likely to have more similar weights, whereas if TUs are bunched up together in 1D, these are much more likely to gather into the same cluster, which becomes dominant in terms of the topological weights, bringing down the structural diversity. 
To understand this quantitatively, consider the topologies with a single tie for which $Z_{\cal G}
\sim Z_i \sim l_i^{3/2}$. In the monodisperse case, when $l_i$ are all equal, all networks with one tie have the same 
weight. All networks with two ties will also have equal weights as $Z_{\cal G} \sim [l_i l_j/(l_i+l_j)]^{3/2}$.

The result in Fig.~\ref{fig:structuraldiversity02} and the associated physical interpretation may be of biophysical relevance, as gene topoi with the same number of TUs can have widely different structural diversity, as shown in~\cite{Chiang2024}~\footnote{The reader should note that in Ref.~\cite{Chiang2024} the definition of structural diversity, though related, was slightly different from the one used here, as it was given in terms of promoter-centred networks, or in other words sets of TUs that form a loop with the promoter.}. Because the positioning of TUs is different in different gene topoi, we suggest that this positioning is at the heart of the observed difference in structural diversity. 

\subsection{Two-cluster configurations with singletons}

In the networks considered in the previous example, there were no singletons, or lone TUs. In other words, every TU participated in a cluster. It is also interesting to study networks in which there are singletons, and in this Section we compute the structural diversity for this special case in this class.  This is more realistic in terms of the application to chromatin loop networks, because polymerases are limited in reality and not all TU will be bound to a transcription factor or polymerase, and if this is the case, then those TUs are unlikely to be in cluster, because they would not participate in bridging-induced phase separation~\cite{Brackley2021}. 

For the case of a network with $n$ TUs, $n_c$ clusters and a fixed number $n_s$ of singletons, there are  
\begin{equation}
    N(n,n_c,n_s) = \binom{n}{n_s} N(n-n_s,n_c) 
\end{equation}
labelled networks that can be formed, where $N(n,n_c)$ is the number of networks with $n$ TUs and
$n_c$ clusters without singletons. 
Also networks with an arbitrary (and not fixed) number of singletons can be counted~\cite{Bonato2024}, however, as we want to compare topological weights it is in practice simpler to restrict to the case of fixed $n_s$, as these networks, in the ``tight graph representation'' of Section III, have the same number of delta functions appearing in $\delta({\mathcal G})$, hence their topological weights have the same dimensions and can be readily compared, without the need to introduce additional normalisation factors. For concreteness, we consider the case of $n=10$, $n_c=2$ and $n_s=2$ [see Fig.~\ref{fig:structuraldiversity02}], which is related to the case of $n=8$, $n_c=2$ and $n_s=0$ considered in the previous section. Note that the total number of possible labelled networks in this worked example is $N(10,2,2)=45 N(8,2) = 5355$.

\begin{figure}
    \centering
    \includegraphics[width=1\linewidth]{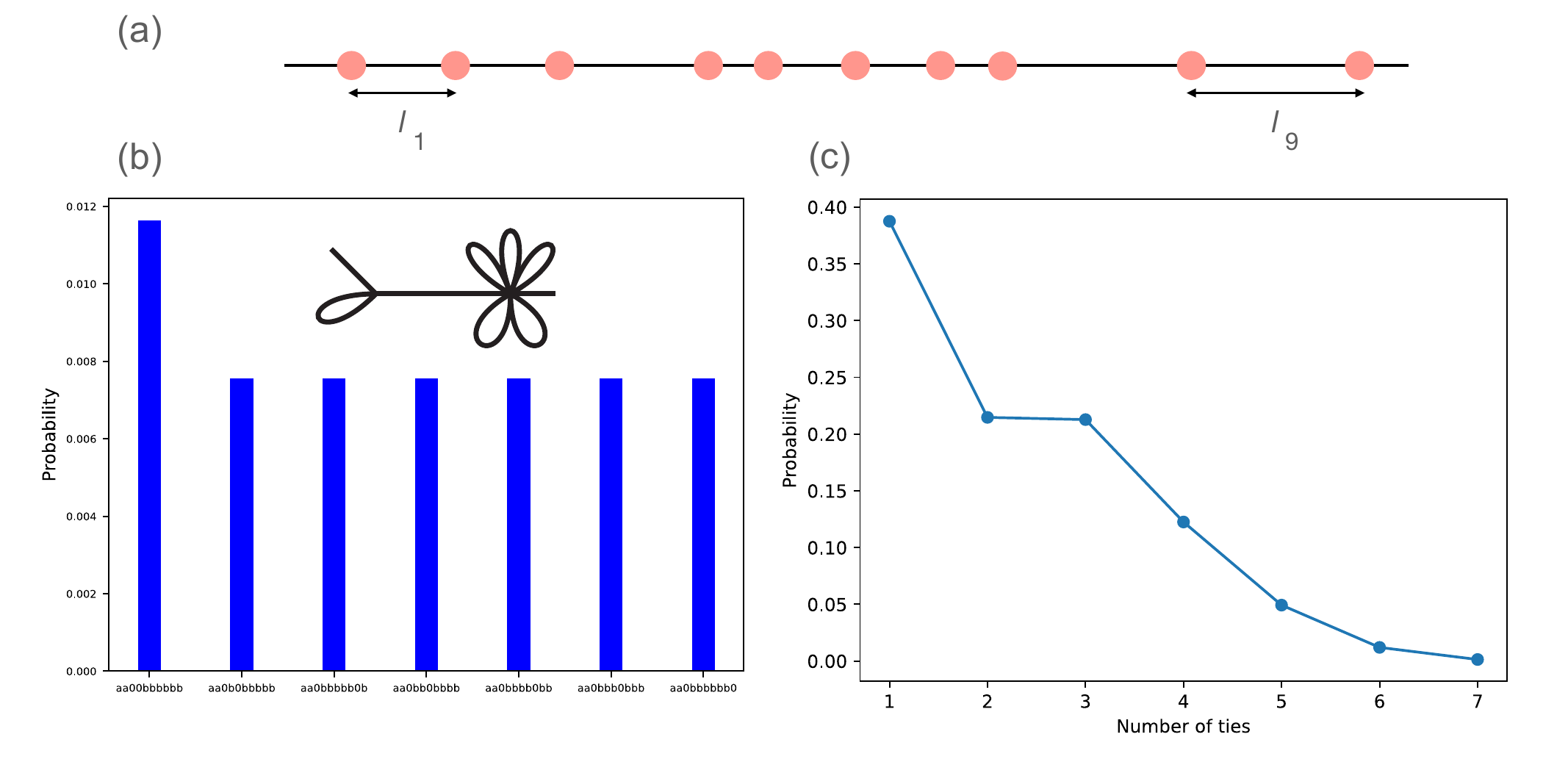}
    \caption{(a) We consider a chromatin fibre with $n=10$ transcription units, which have variable spacing between them. The sketch is a generic example. Configurations are required to have two clusters and exactly two singletons for simplicity. (b) Curve showing the probability, or relative labelled topological weight with respect to the sum of the weights, for the system in (a) with $\{l_i\}_{i=1,\ldots,9}=\{5,34,11,15,6,3,18,4,6\}$, which gives representative results of a generic random disposition of TUs, with Poisson distribution.
    (c) Probability of configurations with a given number of ties $n_t$. }
    \label{fig:structuraldiversity03}
\end{figure}

Fig.~\ref{fig:structuraldiversity03}(b,c) shows the top 5 labelled topologies and the probability of forming a generic network with $n_t$ ties for the case under consideration. The results are qualitatively similar to the case without singletons, in that rosette topologies without ties dominate. The probability of formation of any given single topology is now much lower, because there are $\binom{n}{n_s}$ more of these, but, interestingly, all the top $5$ topologies correspond to the same unlabelled topologies (so singletons decorate the same topology in a different way). The structural diversity in this case is ${\mathcal{S}\simeq 7.55}$, whereas the maximal Shannon entropy, for an equally likely set of topologies with $(n,n_c,n_s)=(10,8,2)$, is $\log{5355}\simeq 8.59$.

\section{Discussion and conclusions}

In summary, in this work we have presented a theoretical framework to compute the combinatorial multiplicity and topological weight of polymer loop networks with arbitrary distances between the nodes, or ``polydisperse loop networks''. Our work is motivated by the fact that a chromatin fibre with transcription units (TUs), which are brought into contact with each other by multivalent complexes, spontaneously forms (through bridging-induced phase separation~\cite{Brackley2013,Brackley2016,Brackley2021,Ryu2021}) polydisperse chromatin loop networks which can be studied with this theory.  Our calculation generalises the previous work in~\cite{Bonato2024}, which was restricted to the monodisperse case, where the distance between consecutive TUs was equal -- in other words, where TUs were uniformly spaced within the chromatin fibre. 

The topological weights we compute are useful to determine the relative frequencies with which different structures arise as genes (or gene topoi~\cite{Chiang2024}) fold in 3D. For labelled loop networks, where different TUs in a topology are labelled, we show that the computation of the topological weight of any given network can be done explicitly, provided we assume that the network is Gaussian, or equivalently, we neglect self- and mutual-avoidance of the polymer segments in the network. The explicit calculation shows there is an intriguing analogy with the computation of the effective resistance of an associated network of Kirchhoff resistors. The results confirm the previous finding obtained for monodisperse chromatin loop networks, that rosette-like topologies, with local loops (connecting a TU to its nearest neighbours along the chain) vastly outnumber non-local loops, have a larger (and usually much larger) weight with respect to more complex networks with non-local wiring between distant TUs along the chain. This is consistent with the numerical finding that rosette-like topologies dominate the topological spectra of 3D gene folding~\cite{short}.

In eukaryotic chromosomes, different transcription units are fundamentally distinct, as they correspond to different genomic sequences associated with specific regulatory elements such as promoters or enhancers. Therefore, labelled networks are more directly related to real chromatin fibres. However, it is also of interest to classify different types of local chromatin folding according to their corresponding unlabelled loop network, much in the same way as it is of interest to classify polymer loops into different knot types. Finding the number of unlabelled loop networks with a fixed number of clusters $n_c$ is a hard problem, which, to our knowledge, has been solved explicitly only for very small values of $n_c$. Quite remarkably, instead, finding the number of labelled loop networks which correspond to any given unlabelled network -- which we refer to as the combinatorial multiplicity of the latter -- is a non-trivial, but solvable, problem. We have shown in Section IV that the problem can be solved by first gluing the entry and exit nodes of the network together, to obtain an Eulerian graph, and then using the BEST theorem of combinatorics to count the number of traversals corresponding to that graph. Combining the topological weights of all labelled networks corresponding to an unlabelled topology, we instead find the topological weight of that topology, which determines the likeliness with which it would be observed in a chromatin fibre corresponding to a gene topos (a promoter of a gene with its contacting TU partners). Comparing the topological weight and combinatorial multiplicity of a topology then allows us to determine, for instance, the relative contribution of combinatorial choices and configurational entropy in determining the relevance, or likeliness of occurrence, of a given network in practice. 

Whilst the theoretical framework we have developed can be used to address a number of properties of chromatin loop networks, in this work, we have applied it to find the structural diversity of a toy model for a gene topos. We have quantified here the structural diversity as the Shannon entropy of a set of chromatin loop networks, which correspond to the possible types of folding of a chromatin fibre with $n$ TUs associated with a gene topos. Because our theory allows for arbitrary patterns of 1D positioning of TUs along the chromatin fibre, we can see how TU disposition affects structural diversity. In turn, the Shannon entropy of a gene folding in 3D is thought to be associated with transcriptional noise, or with the variability in the transcriptional activity of that gene~\cite{Chiang2024,Chiang2025}: therefore, our calculation provides a very simple way to predict transcriptional noise starting only with the 1D positions of TUs around a promoter. We find that structural diversity is maximal for uniformly spaced TUs, whereas it decreases for 1D positioning patterns with larger variability. This is because if TUs are uniformly spaced, there is less difference between the weights of different topologies, whereas if some TUs are close together in 1D, the topologies in which they are in the same cluster have a larger topological weight, thereby diminishing diversity. 
It would be of interest to test this prediction with simulations of mammalian chromatin folding~\cite{Chiang2024}, or with single-cell RNA-seq~\cite{Sukys2024} or single-cell EU-seq~\cite{Xu2024} (nascent transcription) experiments.

We anticipate that our framework could be generalised to compute the probability distribution function for the distance between arbitrary points in a chromatin loop network, which could be used to compare with the non-Gaussian FISH distributions found experimentally~\cite{Remini2024,Remini2025}. Additionally, while we have imagined here that the chromatin loop networks we analyse arise from bridging-induced phase separation, and have TUs as nodes, cohesin-CTCF loops and loop extrusion~\cite{Fudenberg2016,Hansen2017} will also form loop networks~\cite{Polovnikov2023}, whose topology can be studied with similar methods, although in that case non-equilibrium effects will likely need to be explicitly accounted for. We hope that these problems, and others that build on the framework described here, can be studied in the future. 

\section*{Acknowledgements}

 This work was supported by the Wellcome Trust (223097/Z/21/Z). The work of S.K. was supported by Leverhulme 
 Research Fellowship (RF-2023-065\textbackslash 9). E.O. acknowledges support from grant PRIN 2022R8YXMR funded 
 by the Italian Ministry of University and Research. D.M. is grateful to the Francqui Foundation (Belgium) for
 financial support, and to the Department of Physics and Astronomy of KU Leuven, where part of this work was 
 done, for kind hospitality. For open access, the authors have applied a Creative Commons Attribution (CC BY) 
 licence to any Author Accepted Manuscript version arising from this submission. 

\bibliographystyle{apsrev4-1}
\bibliography{references_chromatin_kirchhoff}

\appendix

\section{Combinatorial multiplicities of three-cluster configurations}

In this Appendix, we show how to enumerate all possible three-cluster configurations. 

Up to symmetry, there are six possible cases for configurations with three clusters, and they are listed in Table~\ref{3-clusters-all-cases}. Note that four of these cases are chain-like, and hence have already been considered in Section~\ref{chain-conf-3-clust-sec}. The remaining two cases, configurations 5 and 6 in Table~\ref{3-clusters-all-cases}, correspond to qualitatively different cases, and to triangular defect networks. Their combinatorial multiplicity can still be computed with the BEST theorem, but the calculations are more complicated, and they are outlined in this Appendix in the two following Sections.  

\begin{table}
\begin{center}
\begin{tabular}{c|c|c|c}
 & Configuration & \# edges $n$ & Formula \\
\hline
1. & \begin{minipage}{0.3\textwidth} \begin{tikzpicture}[node distance=1cm,auto,main node/.style={fill,circle,draw,inner sep=0pt,minimum size=3pt}]

\node (1) {};
\node (2)[main node] [right of=1,xshift=-0.3cm] {};
\node (3)[main node] [right of=2,xshift=0.2cm] {};
\node (4)[main node] [right of=3,xshift=0.2cm] {};
\node (5)  [right of=4,xshift=-0.3cm] {};

\path
(1) edge (5);

\path
(2) [loop above] edge [in=120,out=60,loop] (2)
(3) [loop above] edge [in=120,out=60,loop] (3)
(4) [loop above] edge [in=120,out=60,loop] (4);

\node [above of=2, yshift=-0.1cm] {{\tiny $\ell_1$}};
\node [above of=3, yshift=-0.1cm] {{\tiny $\ell_2$}};
\node [above of=4, yshift=-0.1cm] {{\tiny $\ell_3$}};

\node [above of=2, yshift=-0.85cm,xshift=0.6cm] {{\tiny $2k_1+1$}};
\node [above of=3, yshift=-0.85cm,xshift=0.6cm] {{\tiny $2k_2+1$}};

\path
(1) [->,>=stealth] edge (2)
(4) [->,>=stealth] edge (5);

 \end{tikzpicture} \end{minipage}& $\ell_1+\ell_2+\ell_3+2(k_1+k_2)+3$ & (\ref{3-chain-1}) \\

\hline

2. & \begin{minipage}{0.3\textwidth} \begin{tikzpicture}[node distance=1cm,auto,main node/.style={fill,circle,draw,inner sep=0pt,minimum size=3pt}]

\node (1) {};
\node (2)[main node] [right of=1,xshift=-0.3cm] {};
\node (3)[main node] [right of=2,xshift=0.2cm] {};
\node (4)[main node] [right of=3,xshift=0.2cm] {};
\node (5)  [below of=3,yshift=0.5cm] {};

\path
(1) edge (4);

\path
(2) [loop above] edge [in=120,out=60,loop] (2)
(3) [loop above] edge [in=120,out=60,loop] (3)
(4) [loop above] edge [in=120,out=60,loop] (4);

\node [above of=2, yshift=-0.1cm] {{\tiny $\ell_1$}};
\node [above of=3, yshift=-0.1cm] {{\tiny $\ell_2$}};
\node [above of=4, yshift=-0.1cm] {{\tiny $\ell_3$}};

\node [above of=2, yshift=-0.85cm,xshift=0.6cm] {{\tiny $2k_1+1$}};
\node [above of=3, yshift=-0.85cm,xshift=0.6cm] {{\tiny $2k_2$}};

\path
(1) [->,>=stealth] edge (2)
(3) [->,>=stealth] edge (5);

 \end{tikzpicture}  \end{minipage} &  $\ell_1+\ell_2+\ell_3+2(k_1+k_2)+2$ & (\ref{3-chain-2})  \\
\hline
3. & \begin{minipage}{0.3\textwidth} \begin{tikzpicture}[node distance=1cm,auto,main node/.style={fill,circle,draw,inner sep=0pt,minimum size=3pt}]

\node (1) {};
\node (2)[main node] [right of=1,xshift=-0.3cm] {};
\node (3)[main node] [right of=2,xshift=0.2cm] {};
\node (4)[main node] [right of=3,xshift=0.2cm] {};
\node (5)  [below of=2,yshift=0.5cm] {};

\path
(1) edge (4);

\path
(2) [loop above] edge [in=120,out=60,loop] (2)
(3) [loop above] edge [in=120,out=60,loop] (3)
(4) [loop above] edge [in=120,out=60,loop] (4);

\node [above of=2, yshift=-0.1cm] {{\tiny $\ell_1$}};
\node [above of=3, yshift=-0.1cm] {{\tiny $\ell_2$}};
\node [above of=4, yshift=-0.1cm] {{\tiny $\ell_3$}};

\node [above of=2, yshift=-0.85cm,xshift=0.6cm] {{\tiny $2k_1$}};
\node [above of=3, yshift=-0.85cm,xshift=0.6cm] {{\tiny $2k_2$}};

\path
(1) [->,>=stealth] edge (2)
(2) [->,>=stealth] edge (5);

 \end{tikzpicture}  \end{minipage} & $\ell_1+\ell_2+\ell_3+2(k_1+k_2)+1$  & (\ref{3-chain-3}) \\
\hline
4. & \begin{minipage}{0.3\textwidth} \begin{tikzpicture}[node distance=1cm,auto,main node/.style={fill,circle,draw,inner sep=0pt,minimum size=3pt}]

\node (1) {};
\node (2)[main node] [above left of=1,xshift=-0.3cm] {};
\node (3)[main node] [right of=2,xshift=0.2cm] {};
\node (4)[main node] [right of=3,xshift=0.2cm] {};
\node (5)  [below of=3,yshift=0.3cm,xshift=0.3cm] {};

\path
(2) edge (4);

\path
(2) [loop above] edge [in=120,out=60,loop] (2)
(3) [loop above] edge [in=120,out=60,loop] (3)
(4) [loop above] edge [in=120,out=60,loop] (4);

\node [above of=2, yshift=-0.1cm] {{\tiny $\ell_1$}};
\node [above of=3, yshift=-0.1cm] {{\tiny $\ell_2$}};
\node [above of=4, yshift=-0.1cm] {{\tiny $\ell_3$}};

\node [above of=2, yshift=-0.85cm,xshift=0.6cm] {{\tiny $2k_1$}};
\node [above of=3, yshift=-0.85cm,xshift=0.6cm] {{\tiny $2k_2$}};

\path
(1) [->,>=stealth] edge (3)
(3) [->,>=stealth] edge (5);

 \end{tikzpicture}   \end{minipage} & $\ell_1+\ell_2+\ell_3+2(k_1+k_2)+1$ & (\ref{3-chain-4}) \\
\hline
5. & \begin{minipage}{0.3\textwidth} \begin{tikzpicture}[node distance=1cm,auto,main node/.style={fill,circle,draw,inner sep=0pt,minimum size=3pt}]

\node (1) {};
\node (2)[main node] [right of=1,xshift=-0.3cm] {};
\node (3)[main node] [above right of=2,xshift=0.2cm] {};
\node (4)[main node] [below right of=3,xshift=0.2cm] {};
\node (5)  [below of=2,yshift=0.5cm] {};

\path
(1) edge (4)
(2) edge (3)
(3) edge (4);

\path
(2) [loop above] edge [in=120,out=60,loop] (2)
(3) [loop above] edge [in=120,out=60,loop] (3)
(4) [loop above] edge [in=120,out=60,loop] (4);

\node [above of=2, yshift=-0.1cm] {{\tiny $\ell_1$}};
\node [above of=3, yshift=-0.1cm] {{\tiny $\ell_2$}};
\node [above of=4, yshift=-0.1cm] {{\tiny $\ell_3$}};

\node [above of=2, yshift=-0.35cm,xshift=0.5cm] {{\tiny $k_1$}};
\node [below of=3, yshift=0.9cm,xshift=0.45cm] {{\tiny $k_2$}};
\node [below of=3, yshift=0.1cm] {{\tiny $k_3$}};

\path
(1) [->,>=stealth] edge (2)
(2) [->,>=stealth] edge (5);

 \end{tikzpicture}   \end{minipage} & \begin{minipage}{0.4\textwidth} $\ell_1+\ell_2+\ell_3+k_1+k_2+k_3+1$ \\ ($k_1\geq k_3$, $k_1,k_2,k_3$ are all even or all odd, or else no traversal exists)  \end{minipage} & (\ref{3-triangle-5}) \\
\hline
6. & \begin{minipage}{0.3\textwidth} \begin{tikzpicture}[node distance=1cm,auto,main node/.style={fill,circle,draw,inner sep=0pt,minimum size=3pt}]

\node (1) {};
\node (2)[main node] [right of=1,xshift=-0.3cm] {};
\node (3)[main node] [above right of=2,xshift=0.2cm] {};
\node (4)[main node] [below right of=3,xshift=0.2cm] {};
\node (5)  [right of=4,xshift=-0.3cm] {};

\path
(1) edge (4)
(2) edge (3)
(3) edge (4);

\path
(2) [loop above] edge [in=120,out=60,loop] (2)
(3) [loop above] edge [in=120,out=60,loop] (3)
(4) [loop above] edge [in=120,out=60,loop] (4);

\node [above of=2, yshift=-0.1cm] {{\tiny $\ell_1$}};
\node [above of=3, yshift=-0.1cm] {{\tiny $\ell_2$}};
\node [above of=4, yshift=-0.1cm] {{\tiny $\ell_3$}};

\node [above of=2, yshift=-0.35cm,xshift=0.5cm] {{\tiny $k_1$}};
\node [below of=3, yshift=0.9cm,xshift=0.45cm] {{\tiny $k_2$}};
\node [below of=3, yshift=0.1cm] {{\tiny $k_3$}};

\path
(1) [->,>=stealth] edge (2)
(4) [->,>=stealth] edge (5);

 \end{tikzpicture}  \end{minipage} &   \begin{minipage}{0.4\textwidth}  $\ell_1+\ell_2+\ell_3+k_1+k_2+k_3+1$ \\ ($k_1$ and $k_2$ have the same parity which is opposite to $k_3$) \end{minipage} & (\ref{3-triangle-6}) \\
 \hline
\end{tabular}
\end{center}
\caption{All possible inequivalent networks, up to symmetry, for the case of three clusters. In the ``number of edges $n$'', the entrance and exit edges are counted as a single edge}\label{3-clusters-all-cases}
\end{table}

\subsection{Configuration 5 in Table~\ref{3-clusters-all-cases}}
For configuration 5 in Table~\ref{3-clusters-all-cases}, also given to the left in Fig.~\ref{triangle-1}, first note that for at least one traversal to exist we must have $k_1\geq k_3$, and $k_1,k_2,k_3$ are all even or all odd. Next, assume that configuration 5 is oriented as shown schematically to the right in Figure~\ref{triangle-1}. Namely, we assume that $k'_1$ (resp., $k''_1$) edges are oriented from $a$ to $c$ (resp., $c$ to $a$), $k'_2$ (resp., $k''_2$) edges are oriented from $c$ to $d$ (resp., $d$ to $c$), and $k'_3$ (resp., $k''_3$) edges are oriented from $d$ to $a$ (resp., $a$ to $d$). So, $k_1=k'_1+k''_1$, $k_2=k'_2+k''_2$ and $k_3=k'_3+k''_3$, where $0\leq k'_i,k''_i\leq k_i$ for $1,2,3$. Not to overcount symmetric configurations, we assume that if $k_1=k_3$ and $\ell_2=\ell_3$ then $k'_1\geq k''_3$. Finally, note that for at least one traversal in the oriented graph to exist, we must have
\begin{eqnarray}
k'_1+k''_2=k''_1+k'_2 \nonumber \\
k'_2+k''_3=k''_2+k'_3 \nonumber \\
k'_1+k''_3=k''_1+k'_3 \nonumber 
\end{eqnarray} 
Note that typically, there will be more than one non-equivalent orientation and the general formula we are about to obtain is fairly involved. However, for the configurations of our interest with a small number of edges, the formula becomes simple. In general, the number of spanning trees rooted at $a=b$ is $$k''_1\cdot k''_2+k''_1\cdot k'_3 + k'_2\cdot k'_3$$
because only the following subgraphs contribute to this number:
\begin{center}
\scalebox{1.5}{
\begin{tabular}{ccc}
\begin{tikzpicture}[node distance=1cm,auto,main node/.style={fill,circle,draw,inner sep=0pt,minimum size=3pt}]

\node (2)[main node] {};
\node (3)[main node] [right of=2,xshift=0.2cm] {};
\node (4)[main node] [right of=3,xshift=0.2cm] {};
\node (6)  [below of=2,yshift=0.8cm] {{\tiny $a$}};
\node (7)  [below of=3,yshift=0.8cm] {{\tiny $c$}};
\node (8)  [below of=4,yshift=0.8cm] {{\tiny $d$}};

\node [above of=2, yshift=-0.8cm,xshift=0.6cm] {{\tiny $k''_1$}};
\node [above of=3, yshift=-0.8cm,xshift=0.6cm] {{\tiny $k''_2$}};

\path
(3) [->,>=stealth] edge (2)
(4) [->,>=stealth] edge (3);

 \end{tikzpicture}
& 
\begin{tikzpicture}[node distance=1cm,auto,main node/.style={fill,circle,draw,inner sep=0pt,minimum size=3pt}]

\node (2)[main node] {};
\node (3)[main node] [right of=2,xshift=0.2cm] {};
\node (4)[main node] [right of=3,xshift=0.2cm] {};
\node (6)  [below of=2,yshift=0.8cm] {{\tiny $a$}};
\node (7)  [below of=3,yshift=0.8cm] {{\tiny $d$}};
\node (8)  [below of=4,yshift=0.8cm] {{\tiny $c$}};

\node [above of=2, yshift=-0.8cm,xshift=0.6cm] {{\tiny $k'_3$}};
\node [above of=3, yshift=-0.8cm,xshift=0.6cm] {{\tiny $k'_2$}};

\path
(3) [->,>=stealth] edge (2)
(4) [->,>=stealth] edge (3);

 \end{tikzpicture}

& 

\begin{tikzpicture}[node distance=1cm,auto,main node/.style={fill,circle,draw,inner sep=0pt,minimum size=3pt}]

\node (2)[main node] {};
\node (3)[main node] [right of=2,xshift=0.2cm] {};
\node (4)[main node] [right of=3,xshift=0.2cm] {};
\node (6)  [below of=2,yshift=0.8cm] {{\tiny $c$}};
\node (7)  [below of=3,yshift=0.8cm] {{\tiny $a$}};
\node (8)  [below of=4,yshift=0.8cm] {{\tiny $d$}};

\node [above of=2, yshift=-0.8cm,xshift=0.6cm] {{\tiny $k''_1$}};
\node [above of=3, yshift=-0.8cm,xshift=0.6cm] {{\tiny $k'_3$}};

\path
(2) [->,>=stealth] edge (3)
(4) [->,>=stealth] edge (3);

 \end{tikzpicture}

\end{tabular}
}
\end{center}  

\begin{figure}
\begin{center}
\scalebox{1.5}{
\begin{tabular}{cc}
\begin{tikzpicture}[node distance=1cm,auto,main node/.style={fill,circle,draw,inner sep=0pt,minimum size=3pt}]

\node (1) {};
\node (2)[main node] [right of=1,xshift=-0.3cm] {};
\node (3)[main node] [above right of=2,xshift=0.6cm] {};
\node (4)[main node] [below right of=3,xshift=0.6cm] {};
\node (5)  [below of=2,yshift=0.5cm] {};
\node (6)  [below of=2,yshift=0.85cm,xshift=0.4cm] {{\tiny $a=b$}};
\node (7)  [below of=4,yshift=0.8cm] {{\tiny $d$}};
\node (8)  [below of=3,yshift=1.1cm,xshift=-0.2cm] {{\tiny $c$}};

\path
(1) edge (4)
(2) edge (3)
(3) edge (4);

\path
(2) [loop above] edge [in=120,out=60,loop] (2)
(3) [loop above] edge [in=120,out=60,loop] (3)
(4) [loop above] edge [in=120,out=60,loop] (4);

\node [above of=2, yshift=-0.1cm] {{\tiny $\ell_1$}};
\node [above of=3, yshift=-0.1cm] {{\tiny $\ell_2$}};
\node [above of=4, yshift=-0.1cm] {{\tiny $\ell_3$}};

\node [above of=2, yshift=-0.4cm,xshift=0.6cm] {{\tiny $k_1$}};
\node [below of=3, yshift=0.8cm,xshift=0.7cm] {{\tiny $k_2$}};
\node [below of=3, yshift=0.1cm] {{\tiny $k_3$}};

\path
(1) [->,>=stealth] edge (2)
(2) [->,>=stealth] edge (5);

 \end{tikzpicture} 
 
 &
 
 \begin{tikzpicture}[node distance=1cm,auto,main node/.style={fill,circle,draw,inner sep=0pt,minimum size=3pt}]

\node (1) {};
\node (2)[main node] [right of=1,xshift=-0.3cm] {};
\node (3)[main node] [above right of=2,yshift=0.5cm,xshift=0.6cm] {};
\node (4)[main node] [below right of=3,yshift=-0.5cm,xshift=0.6cm] {};
\node (5)  [below of=2,yshift=0.5cm] {};
\node (6)  [below of=2,yshift=0.85cm,xshift=0.1cm] {{\tiny $a=b$}};
\node (7)  [below of=4,yshift=0.8cm] {{\tiny $d$}};
\node (8)  [below of=3,yshift=1.1cm,xshift=-0.2cm] {{\tiny $c$}};

\path
(2) [->,>=stealth, bend left=20] edge (3)
(3) [->,>=stealth, bend left=20] edge (2)
(3) [->,>=stealth, bend left=20] edge (4)
(4) [->,>=stealth, bend left=20] edge (3)
(2) [->,>=stealth, bend left=0] edge (4)
(4) [->,>=stealth, bend left=25] edge (2);

\path
(2) [->,>=stealth, loop above] edge [in=60,out=120,loop] (2)
(3) [->,>=stealth, loop above] edge [in=60,out=120,loop] (3)
(4) [->,>=stealth, loop above] edge [in=60,out=120,loop] (4);

\node [above of=2, yshift=-0.1cm] {{\tiny $\ell_1$}};
\node [above of=3, yshift=-0.1cm] {{\tiny $\ell_2$}};
\node [above of=4, yshift=-0.1cm] {{\tiny $\ell_3$}};

\node [above of=2, yshift=-0.0cm,xshift=0.5cm] {{\tiny $k'_1$}};
\node [above of=2, yshift=-0.6cm,xshift=1cm] {{\tiny $k''_1$}};
\node [below of=3, yshift=0.8cm,xshift=0.7cm] {{\tiny $k'_2$}};
\node [below of=3, yshift=0.3cm,xshift=0.2cm] {{\tiny $k''_2$}};
\node [below of=3, yshift=-0.05cm] {{\tiny $k''_3$}};
\node [below of=3, yshift=-0.7cm] {{\tiny $k'_3$}};

\path
(1) [->,>=stealth] edge (2)
(2) [->,>=stealth] edge (5);

 \end{tikzpicture} 

 \end{tabular}
 } 
\end{center}
\caption{Configuration 5 with three clusters and its orientation.}\label{triangle-1}
\end{figure}

Now, in the oriented version of configuration 5
\begin{eqnarray}
\mbox{outdegree}(a) &=& k'_1+k''_3+\ell_1+1 \nonumber\\
\mbox{outdegree}(c) &=& k''_1+k'_2+\ell_2 \nonumber\\
\mbox{outdegree}(d) &=& k''_2+k'_3 + \ell_3 \nonumber
\end{eqnarray}
and the BEST Theorem, with multiplicities taken into account, gives
\begin{footnotesize}
\begin{equation}\label{3-triangle-5} \sum\frac{(k''_1\cdot k''_2+k''_1\cdot k'_3 + k'_2\cdot k'_3)(k'_1+k''_3+\ell_1)!(k''_1+k'_2+\ell_2-1)!(k''_2+k'_3+\ell_3-1)!}{\ell_1!\ell_2!\ell_3!k'_1!k''_1!k'_2!k''_2!k'_3!k''_3!} \end{equation}
\end{footnotesize}
where the sum is taken over all possible $k'_1$, $k''_1$, $k'_2$, $k''_2$, $k'_3$, $k''_3$ (satisfying all restrictions above including $k'_1\geq k'_3$ if $k_1=k_3$ and $\ell_2=\ell_3$). As an example of application of (\ref{3-triangle-5}), consider the following configuration with $n=8$ $\ell_1=\ell_3=0$, $\ell_2=1$ and $k_1=k_2=k_3=2$ (even though $k_1=k_3$, we have that $\ell_2\neq\ell_3$, so no extra restriction $k'_1\geq k''_3$), which has exactly three non-equivalent orientations, two of which are shown below, the remaining one is obtained from the orientation in the middle by reversing all multiple edges:

\begin{center}
\scalebox{1.5}{
\begin{tabular}{ccc}

\begin{tikzpicture}[node distance=1cm,auto,main node/.style={fill,circle,draw,inner sep=0pt,minimum size=3pt}]

\node (1) {};
\node (2)[main node] [right of=1,xshift=-0.3cm] {};
\node (3)[main node] [above right of=2,yshift=0.5cm,xshift=0.6cm] {};
\node (4)[main node] [below right of=3,yshift=-0.5cm,xshift=0.6cm] {};
\node (5)  [below of=2,yshift=0.5cm] {};
\node (6)  [below of=2,yshift=0.85cm,xshift=0.1cm] {{\tiny $a$}};
\node (7)  [below of=4,yshift=0.8cm] {{\tiny $d$}};
\node (8)  [below of=3,yshift=1.1cm,xshift=-0.2cm] {{\tiny $c$}};

\path
(2) [bend left=20] edge (3)
(3) [bend left=20] edge (2)
(3) [bend left=20] edge (4)
(4) [bend left=20] edge (3)
(2) [bend left=0] edge (4)
(4) [bend left=25] edge (2);

\path
(3) edge [in=60,out=120,loop] (3);

\path
(1) [->,>=stealth] edge (2)
(2) [->,>=stealth] edge (5);

 \end{tikzpicture}

 &
 
\begin{tikzpicture}[node distance=1cm,auto,main node/.style={fill,circle,draw,inner sep=0pt,minimum size=3pt}]

\node (1) {};
\node (2)[main node] [right of=1,xshift=-0.3cm] {};
\node (3)[main node] [above right of=2,yshift=0.5cm,xshift=0.6cm] {};
\node (4)[main node] [below right of=3,yshift=-0.5cm,xshift=0.6cm] {};
\node (5)  [below of=2,yshift=0.5cm] {};
\node (6)  [below of=2,yshift=0.85cm,xshift=0.1cm] {{\tiny $a$}};
\node (7)  [below of=4,yshift=0.8cm] {{\tiny $d$}};
\node (8)  [below of=3,yshift=1.1cm,xshift=-0.2cm] {{\tiny $c$}};

\path
(2) [->,>=stealth, bend left=20] edge (3)
(2) [->,>=stealth, bend right=20] edge (3)
(3) [->,>=stealth, bend left=20] edge (4)
(3) [->,>=stealth, bend right=20] edge (4)
(4) [->,>=stealth, bend left=0] edge (2)
(4) [->,>=stealth, bend left=30] edge (2);

\path
(3)  [->,>=stealth, loop above]  edge [in=60,out=120,loop] (3);

\path
(1) [->,>=stealth] edge (2)
(2) [->,>=stealth] edge (5);

 \end{tikzpicture} 

&

\begin{tikzpicture}[node distance=1cm,auto,main node/.style={fill,circle,draw,inner sep=0pt,minimum size=3pt}]

\node (1) {};
\node (2)[main node] [right of=1,xshift=-0.3cm] {};
\node (3)[main node] [above right of=2,yshift=0.5cm,xshift=0.6cm] {};
\node (4)[main node] [below right of=3,yshift=-0.5cm,xshift=0.6cm] {};
\node (5)  [below of=2,yshift=0.5cm] {};
\node (6)  [below of=2,yshift=0.85cm,xshift=0.1cm] {{\tiny $a$}};
\node (7)  [below of=4,yshift=0.8cm] {{\tiny $d$}};
\node (8)  [below of=3,yshift=1.1cm,xshift=-0.2cm] {{\tiny $c$}};

\path
(2) [->,>=stealth, bend left=20] edge (3)
(3) [->,>=stealth, bend left=20] edge (2)
(3) [->,>=stealth, bend left=20] edge (4)
(4) [->,>=stealth, bend left=20] edge (3)
(2) [->,>=stealth, bend left=0] edge (4)
(4) [->,>=stealth, bend left=25] edge (2);

\path
(3)  [->,>=stealth, loop above]  edge [in=60,out=120,loop] (3);

\path
(1) [->,>=stealth] edge (2)
(2) [->,>=stealth] edge (5);

 \end{tikzpicture} 
 
\end{tabular}
}
\end{center}
In the case of the orientation in the middle, $k'_1=k'_2=k'_3=2$ and $k''_1=k''_2=k''_3=0$. The summand in  (\ref{3-triangle-5}) gives two traversals, which makes sense as the only traversals are $accdacda$ 
and $acdaccda$. 
In the case of the other orientation (above to the right),  $k'_1=k'_2=k'_3=k''_1=k''_2=k''_3=1$ and the formula gives twelve traversals, which makes sense as all traversals are
\begin{center}
\begin{tabular}{cccc}
$accadcda$ & $adccdaca$ & $acadccda$ & $adcdacca$ \\
$adaccdca$ & $accdadca$ & $adacdcca$ & $acdadcca$ \\
$ accdcada$ & $adccacda$ & $acdccada$ & $adcaccda$
\end{tabular}

\end{center}
 
Summing up the numbers of traversals for all three possible orientations (clearly, the third orientation gives the same answer as the one in the middle), we obtain the total number of traversals for the configuration:
$$2+2+12=16.$$
 
\subsection{Configuration 6 in Table~\ref{3-clusters-all-cases}}

\begin{figure}
\begin{center}
\scalebox{1.5}{
\begin{tabular}{cc}
\begin{tikzpicture}[node distance=1cm,auto,main node/.style={fill,circle,draw,inner sep=0pt,minimum size=3pt}]

\node (1) {};
\node (2)[main node] [right of=1,xshift=-0.3cm] {};
\node (3)[main node] [above right of=2,xshift=0.6cm] {};
\node (4)[main node] [below right of=3,xshift=0.6cm] {};
\node (5)  [right of=4,xshift=-0.3cm] {};
\node (6)  [below of=2,yshift=0.85cm,xshift=0.0cm] {{\tiny $a$}};
\node (7)  [below of=4,yshift=0.8cm] {{\tiny $b$}};
\node (8)  [below of=3,yshift=1.1cm,xshift=-0.2cm] {{\tiny $c$}};

\path
(1) edge (4)
(2) edge (3)
(3) edge (4);

\path
(2) [loop above] edge [in=120,out=60,loop] (2)
(3) [loop above] edge [in=120,out=60,loop] (3)
(4) [loop above] edge [in=120,out=60,loop] (4);

\node [above of=2, yshift=-0.1cm] {{\tiny $\ell_1$}};
\node [above of=3, yshift=-0.1cm] {{\tiny $\ell_2$}};
\node [above of=4, yshift=-0.1cm] {{\tiny $\ell_3$}};

\node [above of=2, yshift=-0.4cm,xshift=0.6cm] {{\tiny $k_1$}};
\node [below of=3, yshift=0.8cm,xshift=0.7cm] {{\tiny $k_2$}};
\node [below of=3, yshift=0.1cm] {{\tiny $k_3$}};

\path
(1) [->,>=stealth] edge (2)
(2) [->,>=stealth] edge (5);

 \end{tikzpicture} 
 
 &
 
 \begin{tikzpicture}[node distance=1cm,auto,main node/.style={fill,circle,draw,inner sep=0pt,minimum size=3pt}]

\node (1) {};
\node (2)[main node] [right of=1,xshift=-0.3cm] {};
\node (3)[main node] [above right of=2,yshift=0.5cm,xshift=0.6cm] {};
\node (4)[main node] [below right of=3,yshift=-0.5cm,xshift=0.6cm] {};
\node (5)  [right of=4,xshift=-0.3cm] {};
\node (6)  [below of=2,yshift=0.85cm,xshift=0.0cm] {{\tiny $a$}};
\node (7)  [below of=4,yshift=0.8cm] {{\tiny $b$}};
\node (8)  [below of=3,yshift=1.1cm,xshift=-0.2cm] {{\tiny $c$}};

\path
(2) [->,>=stealth, bend left=20] edge (3)
(3) [->,>=stealth, bend left=20] edge (2)
(3) [->,>=stealth, bend left=20] edge (4)
(4) [->,>=stealth, bend left=20] edge (3)
(2) [->,>=stealth, bend left=0] edge (4)
(4) [->,>=stealth, bend left=25] edge (2);

\path
(2) [->,>=stealth, loop above] edge [in=60,out=120,loop] (2)
(3) [->,>=stealth, loop above] edge [in=60,out=120,loop] (3)
(4) [->,>=stealth, loop above] edge [in=60,out=120,loop] (4);

\node [above of=2, yshift=-0.1cm] {{\tiny $\ell_1$}};
\node [above of=3, yshift=-0.1cm] {{\tiny $\ell_2$}};
\node [above of=4, yshift=-0.1cm] {{\tiny $\ell_3$}};

\node [above of=2, yshift=-0.0cm,xshift=0.5cm] {{\tiny $k'_1$}};
\node [above of=2, yshift=-0.6cm,xshift=1cm] {{\tiny $k''_1$}};
\node [below of=3, yshift=0.8cm,xshift=0.7cm] {{\tiny $k'_2$}};
\node [below of=3, yshift=0.3cm,xshift=0.2cm] {{\tiny $k''_2$}};
\node [below of=3, yshift=-0.05cm] {{\tiny $k''_3$}};
\node [below of=3, yshift=-0.7cm] {{\tiny $k'_3$}};

\path
(1) [->,>=stealth] edge (2)
(2) [->,>=stealth] edge (5);

 \end{tikzpicture} 

 \end{tabular}
 }
\end{center}
\caption{Configuration 6 with three clusters and its orientation}\label{triangle-2}
\end{figure}
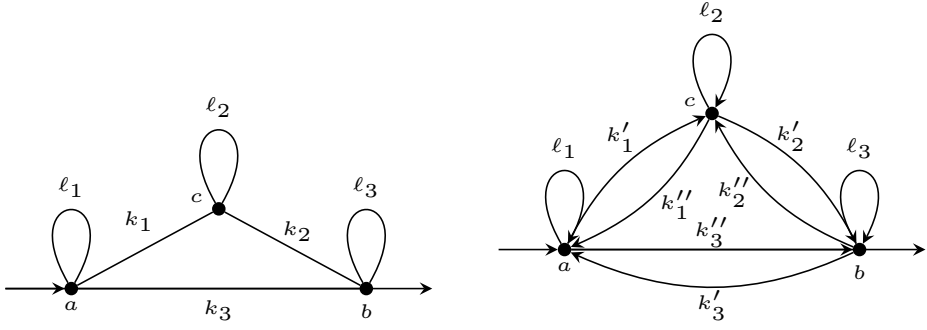

For configuration 6 in Table~\ref{3-clusters-all-cases}, also given to the left in Figure~\ref{triangle-2}, for at least one traversal to exist we must have $k_1$ and $k_2$ are odd (resp., even) and $k_3$ is even (resp., odd), that is, $k_1$ and $k_2$ are of the same parity which is opposite to parity of $k_3$.  Next, similarly to our considerations of configuration 5, assume that configuration 6 is oriented as shown schematically to the right in Figure~\ref{triangle-2}. So, $k_1=k'_1+k''_1$, $k_2=k'_2+k''_2$ and $k_3=k'_3+k''_3$, where $0\leq k'_i,k''_i\leq k_i$ for $1,2,3$. Further, note that for at least one traversal in the oriented graph to exist, we must have
\begin{eqnarray}
k'_1+k''_2=k''_1+k'_2 \nonumber \\
k'_2+k''_3=k''_2+k'_3 \nonumber \\
k'_1+k''_3=k''_1+k'_3 \nonumber 
\end{eqnarray}
which are exactly the same requirements as those for orientations of configuration 5. 

The number of spanning trees rooted at $v$ is $$k'_1\cdot k'_2+k'_2\cdot k''_3 + k''_3\cdot k''_1$$
because only the following subgraphs contribute to this number:
\begin{center}
\scalebox{1.5}{
\begin{tabular}{ccc}
\begin{tikzpicture}[node distance=1cm,auto,main node/.style={fill,circle,draw,inner sep=0pt,minimum size=3pt}]

\node (2)[main node] {};
\node (3)[main node] [right of=2,xshift=0.2cm] {};
\node (4)[main node] [right of=3,xshift=0.2cm] {};
\node (6)  [below of=2,yshift=0.8cm] {{\tiny $a$}};
\node (7)  [below of=3,yshift=0.8cm] {{\tiny $c$}};
\node (8)  [below of=4,yshift=0.8cm] {{\tiny $b$}};

\node [above of=2, yshift=-0.8cm,xshift=0.6cm] {{\tiny $k'_1$}};
\node [above of=3, yshift=-0.8cm,xshift=0.6cm] {{\tiny $k'_2$}};

\path
(2) [->,>=stealth] edge (3)
(3) [->,>=stealth] edge (4);

 \end{tikzpicture}
& 
\begin{tikzpicture}[node distance=1cm,auto,main node/.style={fill,circle,draw,inner sep=0pt,minimum size=3pt}]

\node (2)[main node] {};
\node (3)[main node] [right of=2,xshift=0.2cm] {};
\node (4)[main node] [right of=3,xshift=0.2cm] {};
\node (6)  [below of=2,yshift=0.8cm] {{\tiny $a$}};
\node (7)  [below of=3,yshift=0.8cm] {{\tiny $b$}};
\node (8)  [below of=4,yshift=0.8cm] {{\tiny $c$}};

\node [above of=2, yshift=-0.8cm,xshift=0.6cm] {{\tiny $k''_3$}};
\node [above of=3, yshift=-0.8cm,xshift=0.6cm] {{\tiny $k'_2$}};

\path
(2) [->,>=stealth] edge (3)
(4) [->,>=stealth] edge (3);

 \end{tikzpicture}

& 

\begin{tikzpicture}[node distance=1cm,auto,main node/.style={fill,circle,draw,inner sep=0pt,minimum size=3pt}]

\node (2)[main node] {};
\node (3)[main node] [right of=2,xshift=0.2cm] {};
\node (4)[main node] [right of=3,xshift=0.2cm] {};
\node (6)  [below of=2,yshift=0.8cm] {{\tiny $c$}};
\node (7)  [below of=3,yshift=0.8cm] {{\tiny $a$}};
\node (8)  [below of=4,yshift=0.8cm] {{\tiny $b$}};

\node [above of=2, yshift=-0.8cm,xshift=0.6cm] {{\tiny $k''_1$}};
\node [above of=3, yshift=-0.8cm,xshift=0.6cm] {{\tiny $k''_3$}};

\path
(2) [->,>=stealth] edge (3)
(3) [->,>=stealth] edge (4);

 \end{tikzpicture}

\end{tabular}
}
\end{center}  

Now, in the oriented version of configuration 6
\begin{eqnarray}
\mbox{outdegree}(a) &=& k'_1+k''_3+\ell_1 \nonumber\\
\mbox{outdegree}(c) &=& k''_1+k'_2+\ell_2 \nonumber\\
\mbox{outdegree}(b) &=& k''_2+k'_3 + \ell_3 +1\nonumber
\end{eqnarray}
and the BEST Theorem, with multiplicities taken into account, gives
\begin{footnotesize}
\begin{equation}\label{3-triangle-6} \sum\frac{(k'_1\cdot k'_2+k'_2\cdot k''_3 + k''_3\cdot k''_1)(k'_1+k''_3+\ell_1-1)!(k''_1+k'_2+\ell_2-1)!(k''_2+k'_3+\ell_3)!}{\ell_1!\ell_2!\ell_3!k'_1!k''_1!k'_2!k''_2!k'_3!k''_3!} \end{equation}
\end{footnotesize}
where the sum is taken over all possible $k'_1$, $k''_1$, $k'_2$, $k''_2$, $k'_3$, $k''_3$ (satisfying all restrictions above). As an example of application of (\ref{3-triangle-6}), consider the following configuration with $n=8$ $\ell_1=\ell_3=0$, $\ell_2=k_1=1$, $k_3=2$ and $k_2=3$, which has two non-equivalent orientations:

\begin{center}
\scalebox{1.25}{
\begin{tabular}{ccc}

\begin{tikzpicture}[node distance=1cm,auto,main node/.style={fill,circle,draw,inner sep=0pt,minimum size=3pt}]

\node (1) {};
\node (2)[main node] [right of=1,xshift=-0.3cm] {};
\node (3)[main node] [above right of=2,yshift=0.5cm,xshift=0.6cm] {};
\node (4)[main node] [below right of=3,yshift=-0.5cm,xshift=0.6cm] {};
\node (5)  [right of=4,xshift=-0.3cm] {};
\node (6)  [below of=2,yshift=0.85cm,xshift=0.1cm] {{\tiny $a$}};
\node (7)  [below of=4,yshift=0.8cm] {{\tiny $b$}};
\node (8)  [below of=3,yshift=1.1cm,xshift=-0.2cm] {{\tiny $c$}};

\path
(2)  edge (3)
(3) edge (4)
(3) [bend left=25] edge (4)
(4) [bend left=25] edge (3)
(2) [bend left=0] edge (4)
(4) [bend left=25] edge (2);

\path
(3) edge [in=60,out=120,loop] (3);

\path
(1) [->,>=stealth] edge (2)
(2) [->,>=stealth] edge (5);

 \end{tikzpicture}

 &
 
\begin{tikzpicture}[node distance=1cm,auto,main node/.style={fill,circle,draw,inner sep=0pt,minimum size=3pt}]

\node (1) {};
\node (2)[main node] [right of=1,xshift=-0.3cm] {};
\node (3)[main node] [above right of=2,yshift=0.5cm,xshift=0.6cm] {};
\node (4)[main node] [below right of=3,yshift=-0.5cm,xshift=0.6cm] {};
\node (5)  [right of=4,xshift=-0.3cm] {};
\node (6)  [below of=2,yshift=0.85cm,xshift=0.1cm] {{\tiny $a$}};
\node (7)  [below of=4,yshift=0.8cm] {{\tiny $b$}};
\node (8)  [below of=3,yshift=1.1cm,xshift=-0.2cm] {{\tiny $c$}};

\path
(2)  [->,>=stealth] edge (3)
(4)  [->,>=stealth]  edge (3)
(3) [->,>=stealth, bend left=25] edge (4)
(3) [->,>=stealth, bend right=25] edge (4)
(4) [->,>=stealth, bend left=0] edge (2)
(2) [->,>=stealth, bend right=25] edge (4);

\path
(3)  [->,>=stealth, loop above]  edge [in=60,out=120,loop] (3);

\path
(1) [->,>=stealth] edge (2)
(2) [->,>=stealth] edge (5);

 \end{tikzpicture} 

&

\begin{tikzpicture}[node distance=1cm,auto,main node/.style={fill,circle,draw,inner sep=0pt,minimum size=3pt}]

\node (1) {};
\node (2)[main node] [right of=1,xshift=-0.3cm] {};
\node (3)[main node] [above right of=2,yshift=0.5cm,xshift=0.6cm] {};
\node (4)[main node] [below right of=3,yshift=-0.5cm,xshift=0.6cm] {};
\node (5)  [right of=4,xshift=-0.3cm] {};
\node (6)  [below of=2,yshift=0.85cm,xshift=0.1cm] {{\tiny $a$}};
\node (7)  [below of=4,yshift=0.8cm] {{\tiny $b$}};
\node (8)  [below of=3,yshift=1.1cm,xshift=-0.2cm] {{\tiny $c$}};

\path
(3)  [->,>=stealth] edge (2)
(3)  [->,>=stealth]  edge (4)
(4) [->,>=stealth, bend right=25] edge (3)
(4) [->,>=stealth, bend left=25] edge (3)
(2) [->,>=stealth, bend right=0] edge (4)
(2) [->,>=stealth, bend right=25] edge (4);

\path
(3)  [->,>=stealth, loop above]  edge [in=60,out=120,loop] (3);

\path
(1) [->,>=stealth] edge (2)
(2) [->,>=stealth] edge (5);

 \end{tikzpicture} 
 
\end{tabular}
}
\end{center}
In the case of the orientation in the middle, $k''_1=0$, $k'_1=k'_3=k''_2=k''_3=1$ and $k'_2=2$. The summand in  (\ref{3-triangle-6}) gives eight traversals, which makes sense since in the following four traversals we can swap $cc$ and $c$: $accbcbab$, $accbabcb$, $abaccbcb$ and $abccbacb$. 
In the case of the other orientation (above to the right),  $k'_1=k'_3=0$, $k''_1=k'_2=1$ and $k''_2=k''_3=2$ and the formula gives four traversals, which are $abccbcab$, $abccabcb$, $abcbccab$ and $abcabccb$. 
 
Summing up the numbers of traversals for the two possible orientations, we obtain the total number of traversals for the configuration: $8+4=12$.

\end{document}